\documentclass[prx,notitlepage,superscriptaddress, twocolumn]{revtex4-2}
\usepackage{amsmath}
\usepackage{amssymb}
\usepackage{graphicx}
\usepackage{comment}
\usepackage{wasysym}
\usepackage{cancel}
\usepackage{tikz}
\usepackage{lipsum}

\DeclareUnicodeCharacter{2212}{\ensuremath{-}}

\newcommand{\Ham}{\hat{H}}

\newcommand{\ketbra}[2]{\left|#1\middle>\middle<#2\right|}
\newcommand{\braket}[2]{\left<#1\middle|#2\right>}
\newcommand{\braOPket}[3]{\left<#1\middle|#2\middle|#3\right>}
\newcommand{\bra}[1]{\left<#1\right|}
\newcommand{\ket}[1]{\left|#1\right>}
\newcommand{\OPc}[2]{\hat{#1}_{#2}^{\dag}}
\newcommand{\OP}[2]{\hat{#1}_{#2}^{\vphantom{\dag}}}
\newcommand{\CD}[1]{\OPc{c}{#1}}
\newcommand{\C}[1]{\OP{c}{#1}}
\newcommand{\A}[1]{\OP{a}{#1}}
\newcommand{\AD}[1]{\OPc{a}{#1}}

\newcommand{\E}{\epsilon}

\DeclareMathOperator{\Real}{Re}

\newcommand{\Proj}{\mathcal{P}}

\renewcommand{\k}{\mathbf{k}}
\newcommand{\p}{\mathbf{p}}
\newcommand{\q}{\mathbf{q}}

\newcommand{\ND}[1]{\hat{n}_{#1}}

\newcommand{\captiontitle}[1]{\textbf{#1}}

\newcommand{\Aberry}{\mathcal{A}}

\newcommand{\Jsc}[3]{\hat{\Gamma}^{#1}_{#2\leftarrow#3}}
\newcommand{\Hamfull}{\hat{H}_{f}}
\newcommand{\Hamproj}{\hat{H}_{p}}

\begin{document}
\title{Quantum-Geometric Raman Scattering in Multiorbital Flat-Band Systems}
\date{\today}
\author{Wai Ting Tai}
\email{wttai@sas.upenn.edu}
\affiliation{Department of Physics and Astronomy, University of Pennsylvania, Philadelphia, PA 19104}
\author{Martin Claassen}
\email{claassen@sas.upenn.edu}
\affiliation{Department of Physics and Astronomy, University of Pennsylvania, Philadelphia, PA 19104}

\date{\today}

\begin{abstract}
    Flat-band materials host rich collective phenomena, yet a complete theory of their signatures in inelastic light scattering remains lacking. While naive theories of interacting flat bands would predict that Raman scattering vertices vanish identically in the limit of vanishing dispersion, we show that this picture is incomplete upon including the multiorbital character of such systems. We show that virtual interband processes generate a finite subgap Raman vertex controlled by the quantum geometric tensor even in the strict flat-band limit. We develop a systematic perturbative theory for Raman scattering from flat bands in the limit where the photon energy is far from resonance with interband transitions. Treating interband Coulomb scattering and light-matter coupling on equal footing, we decompose the Raman scattering vertices into an interaction-independent geometric term expressible directly in terms of the quantum geometric tensor, together with effective resonant and non-resonant pieces generated by virtual interband Coulomb scattering.  We then study the polarization-resolved Raman response from collective excitations of an interacting flat band with nontrivial quantum geometry, and demonstrate quantitative agreement of our framework with a full multi-orbital calculation at large photon detuning from interband transitions. These results establish quantum geometry as an intrinsic contribution to inelastic light scattering in flat-band systems, and suggest polarization-resolved Raman spectroscopy as a quantum geometry-sensitive probe of the collective excitations of correlated flat-band platforms.
\end{abstract}

\maketitle

\section{Introduction}

Raman scattering provides frequency-, symmetry-, and polarization-resolved access to the elementary excitations of correlated electron systems \cite{devereaux_inelastic_2007}. The energy loss of inelastically-scattered incident photons provides spectral access to low-energy the collective modes, and polarization rules allow the selection of symmetry-resolved excitations \cite{fleuryScatteringLightOne1968, shastry_theory_1990, khveshchenkoRamanScatteringAnomalous1994, ko_raman_2010}. This flexibility has enabled  studies of elementary excitations in correlated systems, including cuprate superconductors \cite{kleinTheoryRamanScattering1984, devereauxElectronicRamanScattering1994, devereauxElectronicRamanScattering1995, cardonaRamanScatteringHigh1999}, Higgs and Leggett modes in multiorbital superconductors \cite{blumbergObservationLeggettsCollective2007, meassonAmplitudeHiggsMode2014}, spin-phonon coupling in van der Waals magnets \cite{duLatticeDynamicsPhonon2019}, unconventional magnetic excitations \cite{sandilandsScatteringContinuumPossible2015} and charge-density-wave dynamics \cite{kungChiralityDensityWave2015, wuChargeDensityWave2022, liuObservationAnomalousAmplitude2022, kogarLightinducedChargeDensity2020, singhUncoveringHiddenFerroaxial2024}, etc. Theoretically, simplified low-energy descriptions provide useful reference points for Raman scattering. For magnetic excitations in Mott insulators, the Fleury-Loudon theory \cite{fleuryScatteringLightOne1968} describes light coupling to spin degrees of freedom, extended by Shastry and Shraiman via a strong-coupling expansion \cite{shastry_theory_1990}. For parent electronic models, standard theoretical approaches instead invoke the effective-mass description of the Raman vertex \cite{kleinTheoryRamanScattering1984, devereauxElectronicRamanScattering1994, devereauxElectronicRamanScattering1995}, in which the Raman vertex is proportional to the curvature of the intraband energy dispersion. These constructions describe different limits of resonant and non-resonant electronic contributions for the active bands in a solid \cite{devereaux_inelastic_2007}, neither of which however determine the Raman vertex of a multiorbital flat band system.

This problem is most manifest in materials with flat or nearly flat bands near the Fermi level, exemplified by transition-metal oxyhalides \cite{jiaNiobiumOxideDihalides2019,helmerComputationalScreeningMOX22025,luoRobustOrbitalSelectiveFlat2026, baoObservationIsolatedFlat2026} or moir\'e materials  \cite{caoCorrelatedInsulatorBehaviour2018, caoUnconventionalSuperconductivityMagicangle2018, reganOpticalDetectionMott2020, wangCorrelatedElectronicPhases2020, andreiGrapheneBilayersTwist2020,andreiMarvelsMoireMaterials2021}. If the photon energy is not resonant with interband excitations, the conventional dispersive Raman vertex -- which depends on the velocity and curvature of the active band -- is suppressed and vanishes identically in the strict flat-band limit. In these same systems, the role of quantum geometry in governing electromagnetic properties has been increasingly recognized over the past decade \cite{maTopologyGeometryNonlinear2021, tormaEssayWhereCan2023, liuQuantumGeometryCondensed2025, yuQuantumGeometryQuantum2025, jiangRevealingQuantumGeometry2025, vermaQuantumGeometryHidden2026}. It has been shown that geometric contributions enter flat-band superfluid weights \cite{peottaSuperfluidityTopologicallyNontrivial2015, vermaOpticalSpectralWeight2021, tormaSuperconductivitySuperfluidityQuantum2022}, low-frequency optical responses, \cite{souzaPolarizationLocalizationInsulators2000, ghoshProbingQuantumGeometry2024, maoLowenergyOpticalAbsorption2025, carmichaelProbingQuantumGeometry2025, chiuOpticalSignaturesFlat2026}, optical sum rules \cite{onishiQuantumWeightFundamental2024, vermaInstantaneousResponseQuantum2025}, and nonequilibrium and driven dynamics \cite{ozawaExtractingQuantumMetric2018, walickiFloquetEngineeringNearly2024}. Long-lived isolated flat bands have also been proposed in electronic systems coupled to engineered dissipative environments \cite{talkingtonDissipationInducedFlatBands2022}. Quantum geometry has also recently been linked to the Raman response of chiral spin liquids \cite{kollerRamanCircularDichroism2025}. However, the role of quantum geometry in the electronic Raman response of multiorbital flat bands has not been systematically addressed.

Linear optical conductivity provides an instructive contrast: despite the vanishing intraband current vertex in the strict flat-band limit, optical response in flat bands can remain finite via virtual interband scattering processes, governed a combination of Coulomb scattering and Bloch-state geometry \cite{maoDiamagneticResponsePhase2023, tai_quantumgeometric_2023,antebiDrudeWeightInteracting2024,carmichaelProbingQuantumGeometry2025}. However, this does not straightforwardly carry over to inelastic light scattering, as the Raman vertex involves two current insertions connected by a resolvent of the many-body Hamiltonian. It is therefore not self-evident that the same geometric quantities also control the resulting inelastic scattering matrix element.  While inelastic light scattering in free-electron moir\'e systems has been studied \cite{garcia-ruizElectronicRamanScattering2020}, the interband-transition setting differs from the response of an isolated active manifold of interacting flat bands considered here.

To describe the sub-band-gap response of an isolated active manifold, it is instructive to employ a controlled expansion in the inverse separation between the active and remote bands, thereby integrating out the energetically detuned remote bands while retaining information about the orbital admixture of the active bands. In this limit, two effects survive at the same order. Firstly, even when real interband transitions are energetically suppressed, virtual interband scattering exists in a geometric (i.e. adiabatic) sense: as one varies the crystal momentum $\k$ across the Brillouin zone, the Bloch eigenstates continuously rotate within the full multiorbital Hilbert space. This adiabatic evolution is encoded in the quantum geometric tensor (QGT) of the Bloch wavefunctions, comprising the Berry curvature and Fubini-Study metric, which describes the dependence of Bloch wavefunctions under infinitesimal shifts in crystal momentum \cite{provostRiemannianStructureManifolds1980, berryQuantalPhaseFactors1984, restaInsulatingStateMatter2011}. Even for a flat band, the QGT can remain nonzero whenever the Bloch projector depends on momentum. Integrating out virtual interband light-matter processes then generates an effective intraband Raman vertex controlled by this geometry, without requiring real transitions across the remote-band gap.  Secondly, Coulomb interactions can virtually scatter electrons between the active and remote sectors. When combined with interband light-matter vertices, these processes generate interaction-dependent corrections to both the resonant and non-resonant Raman vertices that also survive at the same order. A controlled effective theory must therefore retain the geometric and interaction-assisted contributions on equal footing.

\begin{figure}[t]
    \centering
    \includegraphics[width=0.8\linewidth]{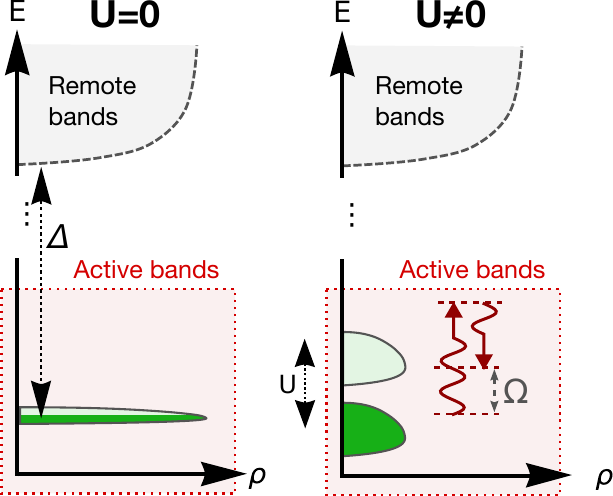}
    \caption{\captiontitle{The quantum-geometric regime of Raman scattering from flatband Mott insulators.} A partially-filled active almost-dispersionless band is separated from remote bands via a band gap $\Delta$, depicted schematically via the local density of states. Coulomb interactions generate an ordered phase. Inelastic (Raman) scattering of photons couples to elementary collective excitations in the active band, with photon energy loss $\Omega \ll \Delta$. The incident photon energy is far detuned from interband excitations.}
    \label{fig:schematic}
\end{figure}

Here, we develop a systematic effective theory of Raman scattering in multiorbital flat-band systems that describes the \textit{emergent} scattering vertices for elementary excitations in an active band manifold (Fig.~\ref{fig:schematic}). The central result is a decomposition of the Raman scattering matrix into three physically distinct contributions: a purely geometric term expressible directly in terms of the QGT of the active band, an effective resonant term with a renormalized current vertex, and an effective non-resonant term generated by virtual interband Coulomb scattering that cannot be determined by the QGT alone. Each contribution survives the flat-band limit at the same parametric order as the intraband velocity term in a conventional dispersive band, rather than as a subleading correction that would be suppressed by interband-transition detuning. We apply the theory to an interacting two-orbital model with a partially-filled flat band, with tunable band gaps and quantum geometry, showing that our effective theory reproduces the full two-band model in both symmetric $A_{\rm 1g}$ and antisymmetric $A_{\rm 2g}$ Raman channels at the limit of large detuning from interband transitions. Meanwhile, when the gap becomes comparable to the relevant photon or many-body energy scales, the active-manifold expansion fails and the full multiband response, including resonant or real interband processes, must be retained. These results establish quantum geometry as an intrinsic contribution to the effective Raman vertex of flat-band systems in both non-interacting and interacting cases, and suggest polarization-resolved Raman spectroscopy as a geometry-sensitive probe in moir\'e Mott insulators.

\section{Review on light-matter coupling and conventional Raman vertex}\label{section:review_raman}

A common theoretical starting point for optical responses of correlated electrons is an effective Hubbard-like description of the active band at the Fermi level. We start by reviewing the theory for inelastic light scattering. The purpose of this section is twofold: (i) to make explicit why a perfectly isolated flat band is optically dark, and (ii) to explain why this picture becomes incomplete in multi-orbital systems, where virtual excitations to remote bands yield effective \textit{intra}band processes that are controlled by quantum geometry. This sets up the central question of this work: how to construct a controlled effective theory of Raman scattering in the presence of multi-orbital entanglement. 

We consider electrons coupled to an electromagnetic field
\begin{align}\label{eq:singleorbitallightmattercouplingHamiltonian}
    \Ham &= \Hamfull
    +\sum_\mu \left( \omega_I \AD{I,\mu} \A{I,\mu} + \omega_F \AD{F,\mu} \A{F,\mu} \right)+ \hat{H}_{\text{lm}},
\end{align}
where $\Hamfull$ denotes the electronic Hamiltonian (including interactions), and $\hat{H}_{\text{lm}}$ is the light--matter coupling. The incoming ($I$) and outgoing ($F$) photons have frequencies $\omega_I$ and $\omega_F$ and polarizations $\mu$, with $\AD{}$ ($\A{}$) the corresponding creation (annihilation) operators. We can then expand $\hat{H}_{\text{lm}}$  to quadratic order in the vector potential $\hat{A}^\mu$, 
\begin{align}\label{eq:V_param_dia}
    \hat{H}_{\text{lm}} = \sum_{\mu} \hat{J}_0^\mu \hat{A}^\mu + \frac{1}{2} \sum_{\mu\nu} \hat{\gamma}_0^{\mu\nu} \hat{A}^\mu \hat{A}^\nu,
\end{align}
with $\hat{J}_0^\mu$ (paramagnetic) and $\hat{\gamma}_0^{\mu\nu}$ (diamagnetic) the electronic current operators \cite{devereaux_inelastic_2007}. Below, spin is assumed to add a trivial Kramers degeneracy and the spin label therefore is suppressed in Bloch wavefunction and geometric quantities unless needed.

Restricting the initial and final photon states to one incoming photon of frequency $\omega_I$ and one outgoing photon of frequency $\omega_F$, the Raman vertex separates into resonant and non-resonant contributions,
\begin{align}\label{eq:ramanscatteringmatrixcontributions}
\hat{M}_{R}^{\mu\nu} &= \Big( \hat{J}_0^\mu \frac{1}{E_0 + \omega_I - \Hamfull} \hat{J}_0^\nu \notag\\
&\hspace{10mm}+ \hat{J}_0^\nu \frac{1}{E_0 - \omega_F - \Hamfull} \hat{J}_0^\mu \Big)\notag~, \\
\hat{M}^{\mu\nu}_{NR} &=\hat{\gamma}_0^{\mu\nu} ~.
\end{align}
The non-resonant contribution corresponds to instantaneous two-photon coupling, while the resonant contribution describes scattering via virtual intermediate electronic states. The scattering intensity is related to these vertices through
\begin{align}\label{eq:ramanscatteringcrosssection}
S(\Omega) = \sum_{F} \left| \braOPket{F}{e^{F*}_{\mu} \hat{M}^{\mu\nu} e^I_{\nu}}{I} \right|^2
\delta(E_F - E_I - \Omega),
\end{align}
with $\hat{M}^{\mu\nu} = \hat{M}_{R}^{\mu\nu} + \hat{M}^{\mu\nu}_{NR}$, $e^{I(F)}$ denoting the polarization vectors of the incoming (outgoing) photon and $\Omega = \omega_I - \omega_F$ is the Raman shift. Here, $S$ contains pure resonant, pure non-resonant, and interference (mixed) contributions. Specific polarization geometries project onto irreducible representations of the crystal point group, defining the symmetry channels of the Raman response \cite{devereaux_inelastic_2007}. For example, for the square lattice, $\hat{M}_{A_{1g}} \propto \hat{M}^{xx} + \hat{M}^{yy}$ and $\hat{M}_{A_{2g}} \propto \hat{M}^{xy} - \hat{M}^{yx}$.

\subsection{Single-orbital case}

For a single-orbital Hubbard model of an isolated band
\begin{align}\label{eq:singleorbitalHamiltonian}
    \Hamfull &= \sum_{\k} \CD{\k} h_{\k}\C{\k}+\hat{H}_{\text{int}}
\end{align}
with dispersion $h_{\k}$ and a density-density interaction $\hat{H}_{\text{int}}$ (e.g. Hubbard interaction), the current vertices reduce to derivatives of $h_{\k}$:
\begin{align}\label{eq:regularcurrentvertex}
	\hat{J}_0^\mu &= \sum_{\k}  \frac{\partial h_{\k}}{\partial k_\mu}\,\CD{\k} \C{\k},\notag  \\
    \hat{\gamma}_0^{\mu\nu} &= \sum_{\k}  \frac{\partial^2 h_{\k}}{\partial k_\mu \partial k_\nu}\,\CD{\k} \C{\k}.
\end{align}
Therefore, in the perfectly flat limit ($h_{\k}=\text{const.}$), both $\hat{J}_0^\mu$ and $\hat{\gamma}_0^{\mu\nu}$ vanish identically. Since $\hat{H}_{\text{int}}$ is unchanged by Peierls substitution, interactions do not restore a conventional single-band current in this limit. This suggests, naively, that an isolated flat band is optically dark.

\subsection{Multi-orbital case}

Accounting for the multiorbital character of materials, the conclusion changes qualitatively for flat bands. The active-band Bloch eigenstate acquires a nontrivial, $k$-dependent orbital composition from its embedding in the larger multiorbital Hilbert space. Consider a general multiorbital electronic Hamiltonian
\begin{align}\label{eq:multiorbitallightmattercouplingHamiltonian}
    \Hamfull &= \sum_{o_1 o_2\k} \CD{o_1\k} h_{o_1 o_2\k} \C{o_2\k} + \hat{H}_{\text{int}}
\end{align}
where $o_1$, $o_2$ index atomic orbitals. Upon transforming to the band basis, the paramagnetic current contains both \textit{intraband} and \textit{interband} components. Writing the band energies as $\epsilon_{n\k}$ and Bloch eigenstates as $\ket{u_{\k n}}$, the paramagnetic current takes the schematic form \cite{blountFormalismsBandTheory1962}
\begin{align}
 \hat{J}_0^\mu &= \underbrace{\sum_{n \k}  \frac{\partial \epsilon_{n\k}}{\partial k_\mu} ~\CD{n\k} \C{n\k}}_{\hat{J}_{0, \text{intraband}}^{\mu}} \\
 & + \underbrace{i \sum_{nm\k} \left( \epsilon_{n\k} - \epsilon_{m\k} \right) \mathcal{A}_{mn}^{\mu}(\k) ~\CD{m\k} \C{n\k}}_{\hat{J}_{0, \text{interband}}^{\mu}}
 \end{align}
where $\mathcal{A}_{mn}^{\mu}(\k) = - i\braket{u_{\k m}}{\partial_{\k_\mu} u_{\k n}}$ is the interband Berry connection. For later convenience we define the interband current operators $\Jsc{\mu}{m}{n}$ via

\begin{align}
    \hat{J}_{0,\mathrm{interband}}^{\mu} &= \sum_{m\neq n} \Jsc{\mu}{m}{n},\\
    \Jsc{\mu}{m}{n} &= i \sum_{\k} \left( \epsilon_{n\k} - \epsilon_{m\k} \right)
    \mathcal{A}_{mn}^{\mu}(\k)\,\CD{m\k} \C{n\k}.\label{eq:transferoperator}
\end{align}
Even when an \emph{active} band is perfectly flat, this interband current operator, together with interband Coulomb interactions, can generate effective intraband Raman vertices that remain finite in the large-gap limit ($\epsilon_{n\k} - \epsilon_{m\k} \gg \omega_I, \omega_F$). Analogous virtual-interband contributions to low-energy electromagnetic response have previously been discussed for driven flat-bands and diamagnetic responses \cite{tai_quantumgeometric_2023,  maoDiamagneticResponsePhase2023}. Section~\ref{section:perturbationscheme} develops the corresponding controlled expansion for Raman processes, explains why these virtual-interband contributions can remain significant even in the large-band-gap limit, and constructs an effective perturbation theory for them.

\section{Perturbation scheme} \label{section:perturbationscheme}

The central result of this paper is a systematic theory of Raman scattering in multi-orbital flat-band systems that describes the emergent coupling of inelastically scattered photons to the low-energy collective excitations of the active bands, in the limit that interband transitions to inert bands are far off resonance. To this end, we devise a downfolding procedure that carefully integrates out inert (energetically off-resonant) bands in favor of effective Raman vertices. We organize this perturbative program in powers of the inverse band gap to remote bands. We treat the intraband interactions within the active manifold exactly and expand perturbatively in the light-matter coupling and interband Coulomb scattering. 

Within each term of this expansion, we then take the large-gap limit, using the photon frequencies and interaction strength to define dimensionless small parameters $\omega_{I,F}/\Delta$ and $U/\Delta$. We retain all contributions that remain finite as $\Delta \rightarrow \infty$ at fixed $\omega_{I,F}$ and $U$, which we denote $O(\Delta^0)$. Corrections are suppressed by at least one power of $\omega_{I,F}/\Delta$ and $U/\Delta$. However, each interband current vertex carries an explicit band-energy prefactor of $O(\Delta)$ [Eq.~(\ref{eq:transferoperator})] , which compensates the $O(1/\Delta)$ energy denominator of one remote-band resolvent. As a result, contributions survive the large-gap limit whenever every remote-band resolvent is paired with an interband current insertion.

Formally, this implies that increasing the separation between the active and remote bands does not by itself force the quantum-geometric contributions to vanish, provided that the system retains appreciable quantum geometry. A consistent theory must therefore carefully integrate out both the interband light-matter coupling and the interband Coulomb interaction at the same order, for dropping either while keeping the other yields an incomplete result at $O(\Delta^0)$. More stringently, projecting out remote bands naively misses contributions that are leading order in the large-gap expansion. Below, we identify a geometric contribution in the non-interacting limit, together with virtual scattering elements that resemble the non-resonant and resonant Raman processes, renormalized by virtual interband interaction.

\subsection{Setup and Power Counting}

We partition the Hilbert space into an active manifold $\mathcal G$ and remote bands separated by a large gap $\Delta$. The Hamiltonian can be partitioned as
\begin{align}
    \Ham &= \Hamproj + \hat{W} +\sum_\mu \left( \omega_I \AD{I,\mu} \A{I,\mu} + \omega_F \AD{F,\mu} \A{F,\mu} \right),
\end{align}
where $\Hamproj$ contains the band-diagonal electronic structure of both active and remote bands together with the intraband interaction within the active manifold,
\begin{align}
    \Hamproj &= \sum_{n \k} \E_{n\k}~ \CD{n\k} \C{n\k} ~+~   \hat{V}^{\{\mathcal{G}\}}_{\text{intraband}},
\end{align}
while $\hat W$ collects all other terms in the Hamiltonian, separated between the \textit{intraband} and \textit{interband} light-matter coupling as well as the \textit{interband interaction terms}
\begin{align}
    \hat{W} =& \sum_{\mu} \hat{J}_{0, \text{interband}}^\mu \hat{A}^\mu  + \sum_{\mu\nu} \hat{\gamma}_{0, \text{interband}}^{\mu\nu} \hat{A}^\mu \hat{A}^\nu \notag\\
    & + \sum_\mu \hat{J}_{0, \text{intraband}}^\mu \hat{A}^\mu + \sum_{\mu\nu} \hat{\gamma}_{0, \text{intraband}}^{\mu\nu} \hat{A}^\mu \hat{A}^\nu  \notag\\
    &+  \hat{V}^{\{\mathcal{G}\}}_{\text{interband}} .
\end{align}

\begin{figure}
    \centering
    \includegraphics[width=\linewidth]{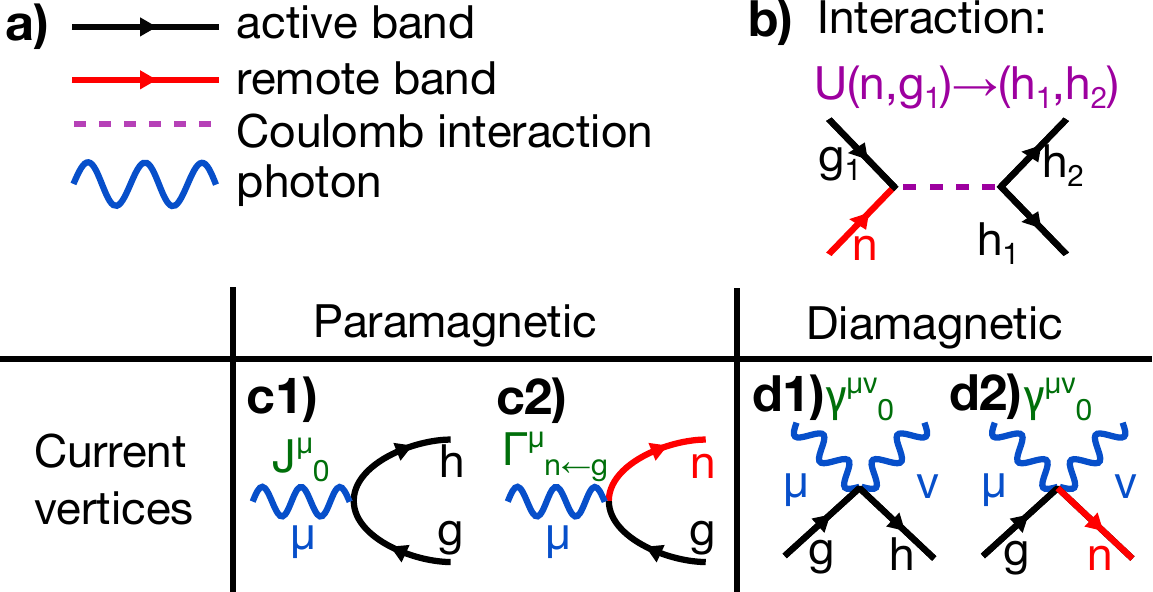}
    \caption{\captiontitle{Diagrammatic representation of the Raman scattering expansion and effective processes.} (a) Legend: black (red) lines denote active band (remote band) electron legs. (b) Interaction vertex $U$ mixing active and remote bands [Eq.~(\ref{eq:interbandinteraction})]. (c-d) Bare paramagnetic and diamagnetic vertices: (c1,d1) intraband and (c2,d2) interband contributions. The intraband paramagnetic current (c1) vanishes in the flat-band limit; (c2) is mediated by $\hat{\Gamma}^\mu_{n\leftarrow g}$ [Eq.~\ref{eq:transferoperator}]. Throughout this and future diagrams, $g$ and $h$ denote active band indices, while $n$ denotes remote band indices.}
    \label{fig:perturbationschematic}
\end{figure}

For concreteness, we use an explicitly on-site Hubbard repulsion
\begin{align}
    \hat V = \frac{U}{2}\sum_i \left(\sum_{o,\sigma} n_{io\sigma}\right)^2
\end{align}
between electrons of different orbital or spin character on the same site. We assume spin-rotation symmetry and suppress spin indices hereafter. Once Fourier transformed into momentum space $\Ham (\k) $ in terms of Bloch wavefunctions  $u^\sigma_{n,o\k}$, the on-site repulsion becomes a momentum-space four-fermion interaction mixing all band indices
\begin{equation}
\begin{split}
	\Ham_{\text{int}} &= \sum_{\substack{n_1n_2n_3n_4}}  V_{n_1n_2\leftarrow n_3n_4}
\end{split}
\end{equation}
where
\begin{equation}\label{eq:V_arrow_def}
\begin{split}
\hat V_{n_1n_2\leftarrow n_3n_4}
&\equiv \frac{U}{L} \sum_{\k\k'\q}\,
\CD{n_1,\k+\q}\CD{n_2,\k'-\q}\C{n_3,\k'}\C{n_4,\k} \\
 & \times \braket{u_{n_1}(\k+\q)}{u_{n_4}(\k)}\braket{u_{n_2}(\k'-\q)}{u_{n_3}(\k')}. 
 \end{split}
\end{equation}
For a longer-range interaction, it suffices to change $U$ to a $\q$-dependent function $U(\q)$. Writing the interaction in band basis allows us to explicitly decompose the interaction into intraband and interband components:
\begin{align}
\hat V^{\{\mathcal G\}}_{\rm intraband}
&\equiv \sum_{g_1g_2g_3g_4\in\mathcal G}\hat V_{g_1g_2\leftarrow g_3g_4},\\
\hat V^{\{\mathcal G\}}_{\rm interband}
&\equiv \sum_{n_1n_2n_3n_4}^{\exists\, n_i\notin\mathcal G}\hat V_{n_1n_2\leftarrow n_3n_4}.\label{eq:interbandinteraction}
\end{align}
The building blocks of this expansion—paramagnetic and diamagnetic vertices, together with the interband interaction—are summarized diagrammatically in Fig.~\ref{fig:perturbationschematic}. Below, we suppress the common Fourier-normalization factor $1/L$ associated with each interaction insertion. The Raman scattering matrix is then expanded by order in the interband perturbation term $\hat{W}$ via the generalized Fermi's golden rule:
\begin{align}\label{eq:generalizedfermigoldenrule}
	M^{\mu\nu}_{FI} &= \left<F\middle| M^{\mu\nu} \middle| I\right>\\
    &= \left<F\middle| \hat{W} + \hat{W} \frac{1}{E_0  + \omega_{ph} -  \Hamproj} \hat{W} + \dots \middle| I\right>,
\end{align}
where we consider a scattering process from $\ket{I}$ to $\ket{F}$ with initial state energy $E_I = E_0$ and explicitly denote the net photon energy absorbed up to the given intermediate state as $\omega_{ph}$: $\omega_{ph} = \omega_I$ after absorption of the incoming photon, $-\omega_F$  after emission of the outgoing photon, and $\omega_I - \omega_F$ after both.

We now organize Eq.~\eqref{eq:generalizedfermigoldenrule} in powers of the inverse remote-band gap $1/\Delta$ (as shorthand for the dimensionless expansion parameter $\max\{\omega_I,\omega_F,U\}/\Delta$).  Each virtual excursion into the remote sector introduces an energy denominator of order $O(\Delta^{-1})$, while each interband paramagnetic current vertex carries an explicit band-energy difference and therefore scales as $O(\Delta)$. Consequently, processes with interband current insertions can survive the large-gap limit: the factors of $\Delta$ from the current vertices compensate the remote-sector propagators. By contrast, additional interband Coulomb vertices do not bring compensating powers of $\Delta$, so terms with too many Coulomb insertions are suppressed by extra powers of $O(\Delta^{-1})$. Throughout, we assume $\omega_I,\omega_F,U\ll\Delta$ and retain all contributions that remain finite as $\Delta\to\infty$.

For Raman scattering, the photon number changes by one incoming and one outgoing photon. Thus, each retained process contains either two paramagnetic current insertions or one diamagnetic insertion, together with any interband Coulomb vertices allowed by the above power counting. We denote by $\hat{M}_i^{\mu\nu}$ the operator-valued contribution containing $i$ insertions of the perturbation $\hat W$, with $E_0$ the ground-state energy entering as a parameter; its matrix elements between eigenstates of $\Hamproj$ enter the golden rule as in Eq.~(\ref{eq:generalizedfermigoldenrule}).  The expansion terminates at fourth order at $O(\Delta^0)$, corresponding to four insertions of $\hat{W}$, as higher-order terms contain additional remote-sector propagators from interband Coulomb scattering, but no additional interband current vertices to compensate them, and are therefore $O(\Delta^{-1})$ or smaller. The Raman matrix can therefore be written as
\begin{align}\label{eq:perturbationtheorydecomposition}
    \hat{M}^{\mu\nu} = \hat{M}^{\mu\nu}_1 + \hat{M}^{\mu\nu}_2 + \hat{M}^{\mu\nu}_3 + \hat{M}^{\mu\nu}_4 + O\left(\frac{\omega_{I/F}}{\Delta}, \frac{U}{\Delta}\right).
\end{align}
The four terms are not all independently meaningful in the large-gap limit. In particular, $\hat{M}^{\mu\nu}_1 + \hat{M}^{\mu\nu}_2$ are the bare non-resonant and resonant contributions, whose  O($\Delta$) contributions cancel exactly, leaving a finite $O(\Delta^0)$ remainder. They can be grouped into three physically distinct contributions to the scattering matrix:
\begin{enumerate}
    \item An effective resonant contribution with a renormalized current vertex arising from $\hat{M}^{\mu\nu}_4$ (two interband currents $+$ two interband Coulomb), that can be expressed in terms of a covariant derivative.
    \item A non-interacting finite geometric residue $\hat{M}^{\mu\nu}_{\text{geom}}$ arising from $\hat{M}^{\mu\nu}_1 + \hat{M}^{\mu\nu}_2$ expressible in terms of the quantum geometric tensor (Appendix \ref{appendix:geometriccontributionderivation}).
    \item A collection of terms frequency-independent at large $\Delta$. Besides the regular intraband diamagnetic term, additional contributions arise from both  $\hat{M}^{\mu\nu}_3$ (two interband currents $+$ one interband Coulomb) and an interaction-driven correction from $\hat{M}^{\mu\nu}_1 + \hat{M}^{\mu\nu}_2$. These terms are frequency-independent and therefore form an effective non-resonant contribution.
\end{enumerate}

These three contributions are derived in the subsections below.

\subsection{Single band}

We now focus on the case where the active manifold $\mathcal G$ consists of a single band, denoted band $0$. (The subscript $0$ on $\hat J_0^\mu$ and $\hat\gamma_0^{\mu\nu}$ still refers to the bare paramagnetic and diamagnetic vertices, not this band index.) Below,  we consider only electron-like scattering processes for simplicity, where an electron is excited from the active band to remote empty bands. Hole-like processes, involving excitation from remote filled bands into the active manifold, can be formulated analogously with filled remote bands replacing empty ones.


\begin{figure}
    \centering
    \includegraphics[width=0.8\linewidth]{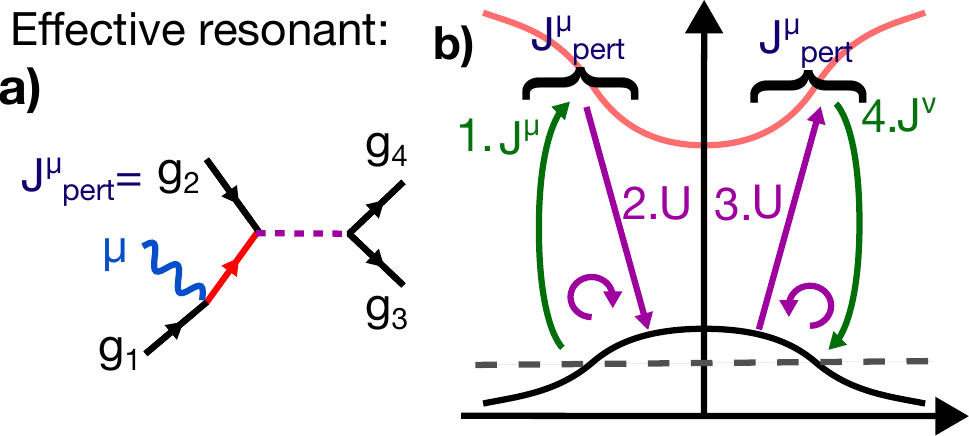}
    \caption{\captiontitle{Schematic of effective resonant term.} (a) Diagrammatic representation of the effective current vertex in Eq.~(\ref{eq:scatteringcurrento3}). The legend is given in Fig.~\ref{fig:perturbationschematic}(a). (b) An equivalent representation of the process from Eq.~(\ref{eq:perturbationm4}), where the intraband current is zero, in terms of the band picture. Here, the perturbation current $J^\mu_{\text{pert}}$ consists of (1) a virtual excitation to the remote band followed by (2) a four-point interband Coulomb scattering.}
    \label{fig:effectiveresonant}
\end{figure}

\subsubsection*{Effective resonant term}

The fourth-order term $\hat{M}^{\mu\nu}_4$ has the form of a resonant contribution with a renormalized current, as shown in Fig.~\ref{fig:effectiveresonant},
\begin{align}\label{eq:perturbationm4}
M_{F,R}^{\mu\nu} &= \bra{F} \Big( \hat{J}_{\text{eff}}^\mu(\omega_I, -\omega_F) \frac{1}{E_0 + \omega_I - \Hamproj} \hat{J}_{\text{eff}}^\nu(0, \omega_I)  \\
&+ \hat{J}_{\text{eff}}^\nu (-\omega_F, \omega_I) \frac{1}{E_0 - \omega_F - \Hamproj} \hat{J}_{\text{eff}}^\mu(0, -\omega_F)  \Big) \ket{I} \notag
\end{align}
where $\hat{J}_{\text{eff}}^\mu(\omega_1, \omega_2) = \hat{J}_{0, \text{intraband}}^\mu+\hat{J}_{\text{pert}}^\mu(\omega_1, \omega_2) $ consists of the bare intraband current and an effective contribution $\hat{J}_{\text{pert}}$ at the same order $O(\Delta^0)$, provided that the active band dispersion $D_a$ is small compared to the band gap $D_a \ll \Delta$. The perturbation current takes the form
\begin{align}\label{eq:scatteringcurrento3}
	\hat{J}_{\text{pert}}^\mu(\omega_1, \omega_2)& = \sum_{m \in \text{empty}}  \Big[\Jsc{\mu}{0}{m} \frac{1}{E_0 + \omega_1 - \Hamproj} \hat{V}_{m0 \gets 00} \notag \\
    &\hspace{5mm}~+~ \hat{V}_{00 \gets 0m} \frac{1}{E_0 + \omega_1 + \omega_2 - \Hamproj} \Jsc{\mu}{m}{0} \Big] 
\end{align}
involving a virtual interband current followed by a Coulomb scattering that results in an effective intraband two-point process. The interactions are defined in Eq.~(\ref{eq:V_arrow_def}), and the interband current operator in Eq.~(\ref{eq:transferoperator}). In particular, in the flat band limit, the bare intraband current vanishes, so the resonant current solely consists of the  perturbative interband correction $\hat{J}_{\text{pert}}$.

The Bloch state representation of $\hat{J}_{\text{pert}}^\mu(\omega_1, \omega_2)$ can be expressed entirely within the ground state manifold as a gauge-covariant derivative  \cite{Parker2019DiagrammaticSemimetals, maoDiamagneticResponsePhase2023} (derivation in Appendix \ref{appendix:4thorderterm}):
\begin{align}\label{eq:4thordertermblochrepresentation}
    j_{\k\k'\q}^\mu = \frac{1}{2} (\mathcal{D}^\mu_{\k} +   \mathcal{D}^\mu_{\k'})\rho_{\k+\q, \k} \rho_{\k' -\q, \k'},
\end{align}
where
\begin{align}
    \mathcal{D}^\mu_{\k} \rho_{\k+\q,\k} =  \frac{\partial \rho_{\k+\q,\k} }{\partial k_\mu} +  i \left(\mathcal{A}^\mu_{\k+\q} \rho_{\k+\q,\k} - \rho_{\k+\q,\k} \mathcal{A}^\mu_{\k} \right) 
\end{align}
and 
\begin{align}
	\rho_{\k,\k'} = \braket{u_{\k}}{u_{\k'}}. \notag
\end{align}
The covariant derivative $\mathcal{D}^\mu_{\k}$, defined with respect to the Berry connection,   measures how the wavefunction overlap $\rho_{\k,\k'}$ varies across the Brillouin zone, relative to parallel transport by the Berry connection. This establishes that the interaction-driven resonant contribution inherits a geometric structure directly analogous to the non-interacting geometric term.

\subsubsection*{Geometric term}

\begin{figure}
    \centering
    \includegraphics[width=0.6\linewidth]{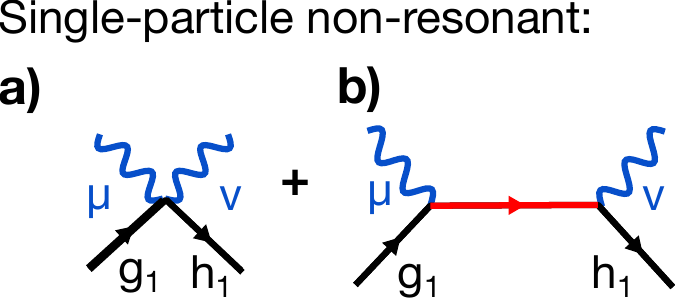}
    \caption{\captiontitle{Schematic of single-particle non-resonant term.} (a-b) The geometric residues represent the non-interacting contributions from (a) the bare non-resonant and (b) the bare resonant terms. Their $O(\Delta)$ contributions cancel out, leaving a non-interacting term of $O(\Delta^0)$ that is given by the \textit{quantum geometric tensor} and an \textit{interacting remainder}.}
    \label{fig:effectivegeometric}
\end{figure}

The leading-order $O(\Delta)$ piece of the projected bare non-resonant vertex $\hat{M}^{\mu\nu}_1$ in Fig.~\ref{fig:effectivegeometric}(a) cancels exactly against the corresponding piece of the bare resonant term $\hat{M}^{\mu\nu}_2$ in Fig.~\ref{fig:effectivegeometric}(b),  built from the non-interacting interband current vertices. Their sum leaves a finite geometric residue plus an interacting correction (see Appendix \ref{appendix:geometriccontributionderivation})
\begin{align}\label{eq:geometrictermandresidue}
\hat{M}^{\mu\nu}_1 +  \hat{M}^{\mu\nu}_2  =&  \hat{M}^{\mu\nu}_{\text{geom}} (\omega_I, \omega_F) + \hat{\gamma}^{\mu\nu}_{\text{residue}}(U) \notag \\
&+ O\left(\Delta^{-1}\right).
\end{align}
Here, $\hat{\gamma}^{\mu\nu}_{\text{residue}}(U)$ denotes the interaction-dependent $O(\Delta^0)$ remainder generated by expanding the resonant denominator, with its explicit form is given in Eq.~(\ref{eq:interactionresidue}) of Appendix~\ref{appendix:geometriccontributionderivation}. The geometric vertex can be expressed directly in terms of the single-band quantum geometric tensor
\begin{align}\label{eq:ramanorder0geometriccontribution}
    \hat{M}^{\mu\nu}_{\text{geom}} &= \sum_{\k} M^{\mu\nu}_{\text{geom}}(\k) \ND{0\k}
    \notag \\
    M^{\mu\nu}_{\text{geom}}(\k)  &= -i \frac{\omega_I + \omega_F}{2} F^{\mu\nu}(\k)-(\omega_I - \omega_F) g^{\mu\nu}(\k),
\end{align}
where $F^{\mu\nu}(\k) = \partial^\mu_k A^\nu(\k) - \partial^\nu_k A^\mu(\k)$ is the Berry curvature and $g^{\mu\nu}(\k) = \Real(\braket{\partial_{k_\mu}u_{0\k}}{\partial_{k_\nu}u_{0\k}} - \braket{\partial_{k_\mu}u_{0\k}}{u_{0\k}}\braket{u_{0\k}}{\partial_{k_\nu}u_{0\k}}) $ is the Fubini-Study metric.  Although $\hat M_{\mathrm{geom}}$ exists already at $U=0$, it is a momentum-weighted occupation operator and commutes with the clean single-band non-interacting Hamiltonian, so it does not connect the ground state to states at different energies. In the interacting problem, $[\hat H_{\text{int}},\hat M_{\mathrm{geom}}]\neq0$ generically, allowing the geometric vertex to generate finite-frequency many-body Raman weight.

\subsubsection*{Two-particle non-resonant term}

\begin{figure}
    \centering
    \includegraphics[width=0.6\linewidth]{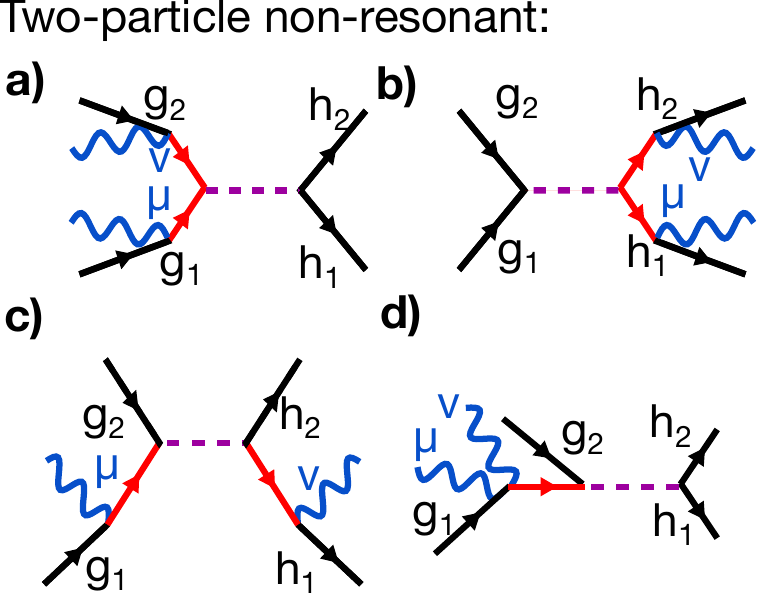}
    \caption{\captiontitle{Schematic of two-particle non-resonant terms}. (a-c) Effective contributions to the non-resonant scattering matrices originating from three-point processes that scatter two electrons, involving two virtual interband current excitations together with an interband Coulomb scattering. (a-c) denote different permutations of the process. (d) A process involving a bare interband diamagnetic current followed by interband Coulomb scattering can also contribute at $O(\Delta^0)$, as the interband diamagnetic matrix element, which contains terms weighted by the active-to-remote energy separation, can scale as $O(\Delta)$. These terms, together with the interacting residue from Fig.~\ref{fig:effectivegeometric}, comprise the effective non-resonant matrix, and in general are functions of the Bloch wavefunctions of the remote bands.}
    \label{fig:effectivenonresonant}
\end{figure}

The non-resonant sector includes all other processes, which are independent of the light frequency at large $\Delta$, which therefore resembles bare non-resonant terms. This includes contributions from three sources at  $O(\Delta^0)$. The first is the interaction correction  $\hat{\gamma}^{\mu\nu}_{\text{residue}}$ from the denominator expansion of $\hat{M}^{\mu\nu}_2$ (Fig.~\ref{fig:effectivegeometric}), which is nonzero only in the presence of interactions (Appendix~\ref{appendix:barenonresonantandresonantcancellation}). The second contribution originates from the third order term $\hat{M}^{\mu\nu}_3$, as shown in Fig.~\ref{fig:effectivenonresonant}(a-c), which involves two interband current vertices dressed with one interband Coulomb scattering (Appendix \ref{appendix:3rdorderterm}). The third contribution is a virtual process including the interband diamagnetic current [Fig.~\ref{fig:effectivenonresonant}(d)]. It contributes at the same order because the interband diamagnetic matrix element contains terms weighted by active-to-remote energy differences of order $\Delta$. Momentum dependence of the remote-band dispersion can provide additional contributions controlled by its bandwidth.

Because each interband current insertion $\Jsc{\mu}{0}{m}\sim O(\Delta)$ compensates one power of the propagator, $\hat{\gamma}^{\mu\nu}_{\text{NR, eff}}$  remains $O(\Delta^0)$ in the large band gap limit. Its frequency independence at $\Delta \rightarrow \infty$ marks it as a \textit{non-resonant} contribution. The Bloch-state representation of one of these terms is worked out in Appendix~\ref{appendix:nonresonantandresonanttermderivation}.  The non-resonant operator itself acts within the active-band manifold, as the fermionic operators create and annihilate electrons in the ground state sector. However, the interaction matrix elements carry an irreducible dependence on remote-band Bloch wavefunctions and dispersions through the exchange channel of the interband Coulomb interaction, so the effective non-resonant term generically cannot be written purely in terms of active-band quantities. Notably, in the $A_{2g}$ channel,  $\hat\gamma^{\mu\nu}_{\text{NR, eff}}$ is symmetric under $\mu\leftrightarrow\nu$ and therefore vanishes identically, leaving only the geometric and resonant terms.

\subsection{Multiband generalization}\label{subsec:multiband}

The extension of the above results to a manifold $\mathcal G$ containing multiple active bands is straightforward. The geometric vertex generalizes to
\begin{align}\label{eq:geomvertex_multiband}
    [\hat{M}^{\mu\nu}_{\text{geom}}]_{g_1 g_2}
      =& \sum_k \CD{g_1\k} \C{g_2 \k}  M^{\mu\nu}_{\text{geom}, g_1 g_2}(\k) \notag\\
      M^{\mu\nu}_{\text{geom}, g_1 g_2}(\k)=& -i\,\frac{\omega_I+\omega_F}{2}\,\mathcal F^{\mu\nu}_{g_1 g_2} (\k) \notag\\
     &- (\omega_I-\omega_F)\,g^{\mu\nu}_{g_1 g_2}(\k), 
\end{align}
where $\mathcal F^{\mu\nu}_{g_1 g_2}(\k)$ and $g^{\mu\nu}_{g_1 g_2}(\k)$ are
the imaginary and real parts of the non-Abelian quantum geometric tensor \cite{wilczekAppearanceGaugeStructure1984, maAbelianNonAbelianQuantum2010}:
\begin{align}
\mathcal Q_{\mu\nu}^{g_1 g_2}(\k)
  &\equiv \braOPket{\partial_{k_\mu}u_{g_1\k}}{(1-P(\k))}{\partial_{k_\nu}u_{g_2\k}}, \\
  \mathcal F^{\mu\nu}_{g_1 g_2}(\k) &= -i  \left[Q_{\mu\nu}^{g_1 g_2}(\k) -  (Q_{\mu\nu}^{g_2 g_1}(\k))^*\right] \\
  g^{\mu\nu}_{g_1 g_2}(\k) &= \frac{1}{2}\left[Q_{\mu\nu}^{g_1 g_2}(\k) +  (Q_{\mu\nu}^{g_2 g_1}(\k))^*\right]
\end{align}
with $P(\k)=\sum_{g\in\mathcal G}\ketbra{u_{g\k}}{u_{g\k}}$ the projector onto the active manifold. The derivation is given in Appendix~\ref{appendix:geometriccontributionderivation}.  The non-resonant and resonant corrections generalize analogously, with the ground state index replaced by band-index sums over $\mathcal G$. In particular, the effective resonant current in Eq. (\ref{eq:4thordertermblochrepresentation}) admits the same multiband generalization, becoming a matrix-valued operator on the active manifold whose covariant derivative is promoted to a non-Abelian form defined by the Berry connection.

\section{Example: two-orbital flat band model}\label{section:twoorbitalflatbandmodel}

\begin{figure}
    \centering
    \includegraphics[width=\linewidth]{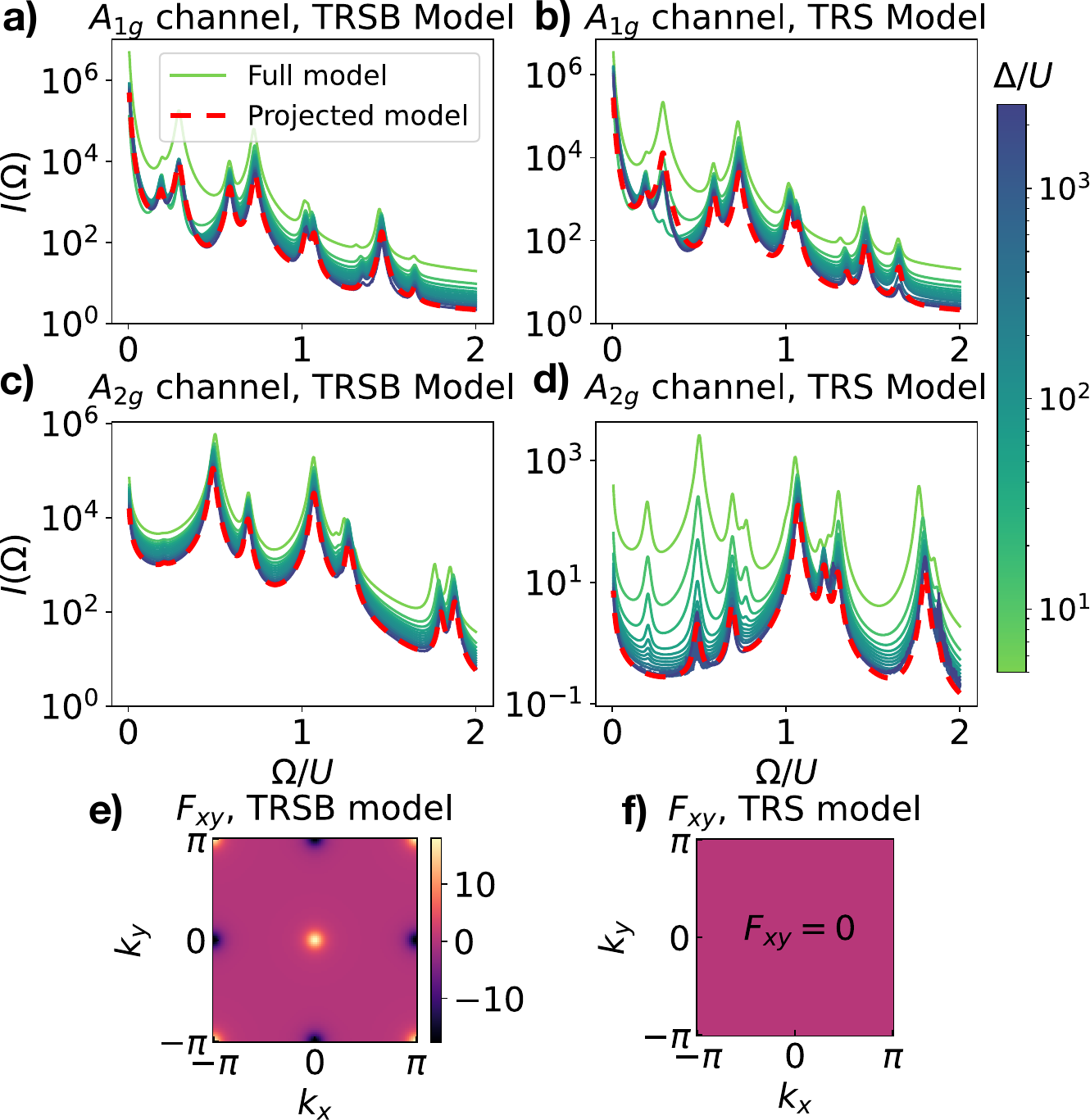}
    \caption{\textbf{Spectral information of the flat band model.} 
    (a-d) Raman spectral response for (a, c) time-reversal symmetry breaking (TRSB) model and (b, d) time-reversal symmetric  (TRS) model defined in Eq. (\ref{eq:flatbandmodeldefinition}) and $A_{1g}$ or $A_{2g}$ channels. The solid lines represent the full model calculations, while the dashed line shows the projected flat-band theory as described in Eq. (\ref{eq:perturbationtheorydecomposition}). Color encodes the ratio of band gap to the interaction strength $\Delta/U$ (see colorbar). The calculation is done at fixed  $U=2.0$, $\omega_I = 4.0$, and dimensionless orbital-texture parameter $t=3.0$. The projected theory is $\Delta$-independent by construction and converges to the exact result for $\Delta \gg U$ and $\Delta \gg \omega_{I}, \omega_{F}$, where $\omega_I$, $\omega_F$ are the frequency of the incoming (I) and outgoing (F) photons. At smaller $\Delta/U$, the effective projected theory does not quantitatively reproduce the full-model Raman intensity because the omitted $O(\Delta^{-1})$ and higher-order corrections remain appreciable, but the two results converge asymptotically as $\Delta/U\to\infty$. (e-f) show the Berry curvature $F_{xy}$  for the models respectively on the Brillouin zone $-\pi \leq kx, ky < \pi$. Since $F_{xy}=0$ for the TRS model, its $A_{2g}$ response originates purely from the effective resonant response in Eq. (\ref{eq:scatteringcurrento3}), while the TRSB model receives contributions from both the geometric response in Eq. (\ref{eq:ramanorder0geometriccontribution}) and resonant response.}
    \label{fig:flatbandsystemresults}
\end{figure}

Our formalism for multi-orbital Raman scattering developed in Section~\ref{section:perturbationscheme} decomposes the Raman vertex into geometric, effective non-resonant, and effective resonant contributions, each surviving the flat-band limit at $O(\Delta^0)$. As a demonstration that the effective theory faithfully captures the full inelastic scattering response from \textit{intra}band excitations, we now study a minimal two-orbital model with a perfectly flat partially-filled interacting active band, using exact diagonalization. While a flat band with strictly-local hopping must have trivial Chern number \cite{chenImpossibilityExactlyFlat2014}, the Berry curvature can be non-zero, leading to observable results from the aforementioned geometric effects. We note, however, that our formalism analogously applies to topological bands. In the flat-band limit, both the intraband conventional intraband velocity vertex and the band-curvature vertex vanish, so all conventional dispersive contributions to the Raman vertex vanish identically. Any Raman signal must therefore originate entirely from the geometric and virtual interband processes derived above, making the magnitude of the Raman response a direct test of orbital admixture and non-quantum geometry.

We start from an interacting two-orbital Hamiltonian of the form
\begin{align}\label{eq:projectorhamiltonian}
  \Ham &= \sum_\k \hat{\mathbf{c}}^\dag_{\k} \mathbf{h}_\k \hat{\mathbf{c}}_{\k} + \frac{U}{2} \sum_{\mathbf{R}} (\hat{n}_{a\mathbf{R}} + \hat{n}_{b\mathbf{R}})^2
\end{align}
where $\hat{\mathbf{c}}_{\k} = (\C{a\k}, \C{b\k})^T$ is the fermion annihilation operator for two orbitals $a$, $b$, and $U$ is an on-site orbitally-isotropic density interaction with number operators $\hat{n}_{a\mathbf{R}}$, $\hat{n}_{b\mathbf{R}}$ where $\C{a\mathbf{R}} = (1/\sqrt{L}) \sum_\k e^{i\k\mathbf{R}} \C{a\k}$. Without loss of generality, the simplest Bloch Hamiltonian hosting a flat isolated zero-energy band with (unnormalized) Bloch state $\tilde{u}_\k$ can be written as a projector
\begin{align}
    \mathbf{h}_\k = \Delta\bigl(\mathcal{N}_\k^2 \mathbf{I} - \ketbra{\tilde{u}_\k}{\tilde{u}_\k}\bigr)
\end{align}
where $\Delta$ is the gap to the second dispersive band, and $\mathcal{N}_\k = \sqrt{\braket{\tilde{u}_\k}{\tilde{u}_\k}}$. If $\tilde{u}_\k$ is analytic, the corresponding tight-binding hopping matrix elements are short-ranged. Because the flat band is dispersionless, the many-body ground state carries no kinetic Raman weight, and interactions within the flat band cannot restore it. Adding a Coulomb interaction can lift the macroscopic degeneracy of the flat-band manifold, selecting a correlated insulating ground state with a well-defined charge gap of order $U$. In addition, it causes the ground state of the full two-band system to acquire components from the remote band, suppressed as $O(U/\Delta)$. In the $\Delta \to \infty$ limit, the ground state is asymptotically confined to the flat-band manifold; however, as argued above, any Raman signal must originate entirely by virtual interband transitions, or equivalently from the non-trivial orbital admixture and geometry of the flat-band Bloch states. Because $\Delta$ enters as an overall energy scale that does not modify the Bloch states $\ket{u_\k}$, tuning $\Delta$ leaves the quantum geometry of the flat band unchanged while controlling the accuracy of the large-gap expansion. This provides a continuous interpolation between the perturbative regime, where the projected theory of Section~\ref{section:perturbationscheme} should be exact, and the small-gap regime, where the expansion loses control. This enables a controlled quantitative test of the projection scheme.

We now study two two-orbital models with (unnormalized) flat-band Bloch wavefunctions
\begin{align}\label{eq:flatbandmodeldefinition}
  \text{(TRSB)}&\quad \ket{\tilde{u}_\k} = 
    \begin{pmatrix} 1 \\ t\,(\sin k_x + i\sin k_y) \end{pmatrix} \notag\\
  \text{(TRS)}&\quad \ket{\tilde{u}_\k} = 
    \begin{pmatrix} 1 \\ t\,(\sin k_x + \sin k_y) \end{pmatrix}.
\end{align}
The first model describes a time-reversal symmetry breaking (TRSB) interacting band with non-trivial Berry curvature; the second model is a time-reversal symmetric (TRS) model with nonzero quantum metric but zero Berry curvature by construction. The band gap scale is set by $\Delta$, while the quantum geometry of the flat band is controlled entirely by the choice of $\ket{\tilde{u}_\k}$. For wavefunctions parameterized by an orbital-texture parameter $t$, the perturbative regime of Section~\ref{section:perturbationscheme}  is controlled by $U/\Delta\ll 1$ and $\omega_{I,F}/\Delta\ll1$, while $t$ determines the quantum geometry held fixed as $\Delta$ is varied.
We emphasize that an unnormalized trigonometric spinor $\ket{\tilde{u}_\k}$ is chosen, so that $\Ham$ has strictly-local hopping in real space. While the lower band is dispersionless, the upper band energy $\Delta\mathcal{N}_\k^2$ has a $\k$-dependent dispersion of order $\Delta$. \footnote{This dispersion also leads to cross terms between interband components of the bare diamagnetic vertex and interband Coulomb scattering, contributing an additional \textit{non-resonant} correction  (Appendix \ref{appendix:barenonresonantcrossterms}).}

Comparing the Raman response from both models captures contributions from both the antisymmetric (Berry curvature) and symmetric (quantum metric) parts of the quantum geometric tensor. The model is solved by exact diagonalization  (see Appendix \ref{appendix:exactdiagonalizationresponse}).  A finite Lorentzian broadening factor $\eta = 0.03$ in the finite-size exact diagonalization calculation is used to produce a continuous spectrum from discrete eigenstates. We note that this can generate remote-band Raman peaks whose Lorentzian tails leak into the low-energy response at a level that does not vanish with increasing $\Delta$. Since this leakage is a numerical artifact, we remove it by projecting the scattering vector onto the active-band manifold before computing the spectral function (Appendix ~\ref{appendix:projectionontogsm}).  The calculation is performed on a $2\times6$ lattice with $N=6$, such that the flat band is half-filled, and in the $K = \Gamma$ momentum sector, with periodic boundary conditions, $t=3.0$. The input frequency $\omega_I = 4.0$ lies  well below the interband gap $\Delta$. The result is shown in Fig.~\ref{fig:flatbandsystemresults}. In the numerical implementation, the effective non-resonant contributions are evaluated using the full multiband Bloch wavefunctions, and therefore retain an explicit dependence on remote-band geometry, consistent with their non-projectable character.

The effective theory of inelastic scattering from the active interacting band -- which is band-gap-independent by construction -- agrees quantitatively with the full model at large detuning from interband transitions $\Delta\gg U,\omega_I,\omega_F$. This agreement is asymptotic, and at smaller $\Delta/U$, the omitted $O(\Delta^{-1})$ and higher-order processes provide corrections to the scattering matrix elements from real interband excitations. Importantly, in the $A_{2g}$ channel, $\hat{\gamma}^{\mu\nu}_{\text{NR, eff}}$ vanishes by symmetry under $\mu\leftrightarrow\nu$, so the response only tracks the effective resonant matrix and the Berry-curvature component of the geometric contribution. By contrast, the $A_{1g}$ channel includes the full set of effective nonresonant contributions. The TRSB model carries nonzero Berry curvature $F_{xy}$, so both the geometric term [Eq.~(\ref{eq:ramanorder0geometriccontribution})] and the effective resonant term [Eq.~(\ref{eq:perturbationm4})] contribute to the $A_{2g}$ response. By contrast, the TRS model has $F_{xy}=0$ by time-reversal symmetry, so the geometric contribution vanishes in $A_{2g}$ and the response arises entirely from the effective resonant term. Quantitative agreement of the effective active-band Raman scattering theory with the full multi-orbital calculation in both cases therefore confirms that the quantum-geometric Raman vertices correctly capture contributions from both the Berry curvature and the interaction-renormalized resonant two-particle currents independently.



\section{Discussion}

We have developed a systematic effective theory of Raman scattering from geometrically-rich or topological flat-band systems. This framework establishes quantum geometry as an intrinsic and interaction-independent contribution to the effective Raman vertex that survives the flat-band limit, in addition to quantum-geometric two-particle scattering vertices. This theory differs from a naive single-band picture, where the conventional dispersive Raman vertex vanishes for a perfectly flat band. The resulting effective vertex separates into a non-interacting geometric contribution that is given by the quantum geometric tensor, an effective non-resonant contribution, and an effective resonant contribution that is determined by a current obtained via a covariant derivative of the Hamiltonian; both latter contributions describe the interaction of a photon with \textit{pairs} of electrons, and can be viewed as an emergent photon dressing of the Coulomb interactions in flat bands. We demonstrate that the resulting theory predicts Raman scattering cross sections for two flatband models with distinct quantum geometries, demonstrating agreement between the full multi-orbital model and the emergent flatband theory.

The most direct extension is to realistic flat-band systems, such as moir\'e transition-metal dichalcogenides \cite{wuTopologicalInsulatorsTwisted2019, devakulMagicTwistedTransition2021, vitaleFlatBandProperties2021, xianRealizationNearlyDispersionless2021, makSemiconductorMoireMaterials2022} and transition-metal oxyhalides \cite{luoRobustOrbitalSelectiveFlat2026, baoObservationIsolatedFlat2026}, which can host flat bands energetically isolated from remote bands. Applied to material-specific correlated Hamiltonians, the formalism developed here allows us to determine which excitations within the active manifold acquire Raman spectral weight and how that weight depends on polarization. This can inform the interpretation of collective modes, such as bimagnons and magnetic excitations, charge excitations in flat-band Mott states, and Raman-active ferroelectric amplitude modes. 

Although our benchmark considers a single isolated active band, the non-Abelian formulation developed above applies to multiband active manifolds. An intriguing extension concerns fractional Chern insulators. Polarization-resolved resonant Raman scattering has been proposed as a probe of magnetoroton modes in fractional quantum Hall fluids \cite{nguyenProbingSpinStructure2021}.  In contrast to conventional Landau levels, moir\'e fractional Chern insulators break Galilean invariance. In particular, FCIs have been predicted to show collective fractional excitons modes which are optically active at the long wavelength limit ($q=0$) \cite{paul2025shining}. Adding external periodic potential can also allow magnetorotons to couple directly to light \cite{kousaTheoryMagnetorotonBands2026}. This motivates asking whether the effective Raman vertex derived here has nonzero matrix elements between an FCI ground state and these collective modes. Evaluating this vertex in an FCI ground state would determine how these collective modes carry observable Raman spectral weight, and how that weight evolves according to quantum geometry fluctuations or violations of FCI ideality conditions.


\section*{Acknowledgments}

The authors thank Benjamin Kass, Deven Carmichael and Spenser Talkington for useful discussions. M. C. and W. Tai were supported by the Department of Energy grant DE-SC0024494. W. Tai was supported by the Croucher Scholarship for Doctoral Study.

\appendix 

\onecolumngrid

\section{Derivation of geometric vertex in a flat band}\label{appendix:geometriccontributionderivation}

This appendix derives  Eq.~(\ref{eq:ramanorder0geometriccontribution}) and its multiband generalization Eq.~(\ref{eq:geomvertex_multiband}). Similar to the main text, we work in the single flat-band case and note the multiband extension at the end of each subsection. We drop the spin index and treat spin as a trivial degeneracy. We take the band gap to be the dominant energy parameter, with both photon frequency and interaction strength much smaller than the gap $U, \omega \ll \Delta$.

\subsection{Non-resonant term from diamagnetic vertex}{\label{appendix:nonresonantdiamagneticexpansion}}

The non-resonant term $\hat{M}^{\mu\nu}_{NR}=\hat\gamma_0^{\mu\nu}$ in Eq.~(\ref{eq:ramanscatteringmatrixcontributions}) is expressed in terms of a second derivative of the orbital Hamiltonian. Expanding in the band basis via the product rule gives
\begin{align}
\frac{\partial^2 h_{o_1 o_2\k}}{\partial k_\mu \partial k_\nu} &=  \frac{\partial^2 u^\dag_{o_1 n}(\k)}{\partial k_\mu \partial k_\nu}\epsilon_{n\k} u_{n o_2}(\k) + u^\dag_{o_1 n}(\k)\frac{\partial^2 \epsilon_{n\k}}{\partial k_\mu \partial k_\nu} u_{no_2}(\k) + u^\dag_{o_1 n}(\k)\epsilon_{n\k}\frac{\partial^2 u_{n o_2}(\k) }{\partial k_\mu \partial k_\nu}  \notag \\
&+ \Big[ \frac{\partial u^\dag_{o_1 n}(\k)}{\partial k_\mu }\frac{\partial \epsilon_{n\k}}{\partial k_\nu} u_{n o_2}(\k) +\frac{\partial u^\dag_{o_1 n}(\k)}{\partial k_\mu }\epsilon_{n\k}\frac{\partial  u_{n o_2}(\k)}{\partial k_\nu} + u^\dag_{o_1 n}(\k) \frac{\partial \epsilon_{n\k}}{\partial k_\mu }\frac{\partial  u_{n o_2}(\k)}{\partial k_\nu} + (\mu\leftrightarrow \nu) \Big].
\end{align}
In the flat band limit ($\epsilon_{0\k} = 0$), projecting onto band $0$ retains only the cross term $\frac{\partial u^*_{o_1 n}(\k)}{\partial k_\mu }\epsilon_{n\k}\frac{\partial  u_{n o_2}(\k)}{\partial k_\nu}  +  (\mu\leftrightarrow \nu)$, with all other terms vanishing. Relabeling in terms of the interband Berry connection and the interband current operators $\Jsc{\mu}{m}{n}$,

 \begin{align}
     \hat{M}^{\mu\nu}_1 = &\sum_{\k o_1 o_2 n}\CD{0 \k} u_{0 o_1} \frac{\partial u^*_{o_1 n}(\k)}{\partial k_\mu }\epsilon_{n\k}\frac{\partial  u_{n o_2}(\k)}{\partial k_\nu} u^*_{o_2 0}\C{0 \k } +(\mu\leftrightarrow \nu) \notag \\
     = &\sum_{n} \Jsc{\mu}{0}{n}\frac{1}{\Ham_0}\Jsc{\nu}{n}{0}+(\mu\leftrightarrow \nu). \notag \\
 \end{align}
This has the same operator structure as a resonant term, with $\Ham_0^{-1}$ playing the role of the energy denominator.  The multiband generalization ($g_1,g_2\in\mathcal G$, $\Ham$ flat) is
\begin{align}
[\hat{M}^{\mu\nu}_1]_{g_1 g_2} = \sum_{n} \Jsc{\mu}{g_1}{n}\frac{1}{\Ham_0}\Jsc{\nu}{n}{g_2}+(\mu\leftrightarrow \nu).
\end{align}

\subsection{Resonant term and cancellation with non-resonant term}\label{appendix:barenonresonantandresonantcancellation}

With the bare intraband current  $\hat J^\mu_{0,\text{intraband}}=0$ vanishing in the flat-band limit, the bare resonant vertex must involve intermediate excitation processes to the excited states and reduces to

\begin{align}\label{eq:fullmodelscatteringcurrento0R}
    \hat{M}^{\mu\nu}_2&=\Jsc{\mu}{0}{n}\frac{1}{E_0 + \omega_I - \Hamproj} \Jsc{\nu}{n}{0} + \Jsc{\nu}{0}{n} \frac{1}{E_0 - \omega_F - \Hamproj} \Jsc{\mu}{n}{0}.
\end{align}
In the large band gap limit, one can expand the denominator. Denoting the intraband interaction term as $H_{U, \text{intraband}}$, and expanding the projected Hamiltonian as $\Ham_p = \Ham_0 + H_{U, \text{intraband}}$,
\begin{align}
    \frac{1}{E_{0} + \omega_I - \Ham_p} &=  - \left( \frac{1}{\Ham_0} + \frac{1}{\Ham_0} (E_0 + \omega_I - H_{U, \text{intraband}})\frac{1}{\Ham_0}  + O(\Delta^{-3}) \right).
\end{align}
The matrix therefore takes the form 
\begin{align}
 \hat{M}_{2}^{\mu\nu} =&\underbrace{-\left(\Jsc{\mu}{0}{n} \frac{1}{\Ham_0} \Jsc{\nu}{n}{0} + \Jsc{\nu}{0}{n} \frac{1}{\Ham_0} \Jsc{\mu}{n}{0}\right)}_{=-\hat{M}^{\mu\nu}_1}  \underbrace{- \left(\Jsc{\mu}{0}{n} \frac{\omega_I}{\Ham^2_0} \Jsc{\nu}{n}{0} - \Jsc{\nu}{0}{n} \frac{\omega_F}{\Ham^2_0} \Jsc{\mu}{n}{0}\right)}_{\equiv \hat{M}^{\mu\nu}_{\text{geom}}} \\
 & \underbrace{- \left(\Jsc{\mu}{0}{n} \frac{1}{\Ham_0} \left(E_0 - H_{U,  \text{intraband}}\right)\frac{1}{\Ham_0}\Jsc{\nu}{n}{0} + \Jsc{\nu}{0}{n}\frac{1}{\Ham_0} \left(E_0 - H_{U,  \text{intraband}}\right)\frac{1}{\Ham_0} \Jsc{\mu}{n}{0}\right)}_{\equiv \hat{\gamma}^{\mu\nu}_{\text{residue}}} + O(\Delta^{-1}),\label{eq:interactionresidue}
 \end{align}
where the truncation now happens at the $O(\Delta^{-1})$ order because $\Jsc{\nu}{0}{n}\sim O(\Delta)$ scales as the band gap. The leading $\hat{M}^{\mu\nu}_2$ term cancels $\hat{M}^{\mu\nu}_1$ exactly when summed. The remaining pieces are $O(\Delta^0)$. The first term, $\hat{M}^{\mu\nu}_{\text{geom}}$ is the geometric residue (simplified below), and the last line gives an interaction correction. Note that in the flat band limit, $E_0$ has no dispersive components and must solely originate from the intraband interaction, so $\hat{\gamma}^{\mu\nu}_{\text{residue}}$ would be zero in the non-interacting case.

The multiband version $\braOPket{g_1}{\hat{M}^{\mu\nu}_2}{g_2}$ with starting state $g_2$ follows identically, with $g_1, g_2\in \mathcal{G}$ replacing the single-band indices:
\begin{align}
     [\hat{M}_{2}^{\mu\nu}]_{g_1 g_2}=&-[\hat{M}_{1}^{\mu\nu}]_{g_1 g_2} - \underbrace{\left(\Jsc{\mu}{g_1}{n} \frac{\omega_I}{\Ham^2_0} \Jsc{\nu}{n}{g_2} - \Jsc{\nu}{g_1}{n} \frac{\omega_F}{\Ham^2_0} \Jsc{\mu}{n}{g_2}\right)}_{\equiv[\hat{M}^{\mu\nu}_{\text{geom}}]_{g_1 g_2}} \\
 &- \underbrace{\left(\Jsc{\mu}{g_1}{n} \frac{1}{\Ham_0} \left(E_{g_2} - H_{U,  \text{intraband}}\right)\frac{1}{\Ham_0} \Jsc{\nu}{n}{g_2} + \Jsc{\nu}{g_1}{n}\frac{1}{\Ham_0} \left(E_{g_2} - H_{U,  \text{intraband}}\right)\frac{1}{\Ham_0} \Jsc{\mu}{n}{g_2}\right)}_{\equiv [\hat{\gamma}^{\mu\nu}_{\text{residue}}]_{g_1 g_2}} + O(\Delta^{-1}).
\end{align}

\subsection{Geometric vertex}\label{appendix:geometricvertexderivation}

This section simplifies $\hat{M}^{\mu\nu}_{\text{geom}}$ in terms of quantum geometric metrics directly from the Berry connection algebra. Expanding out $\Jsc{\mu}{0}{n}=i\sum_\k\Delta_{n0}(\k)\,\mathcal A^{\mu,0n}_\k\CD{0\k}\C{n\k}$,
\begin{align}
\hat{M}^{\mu\nu}_{\text{geom}}=&\sum_{\k} \left[\underbrace{-\Aberry^{\mu, 0n}(\k) \Aberry^{\nu, n0}(\k) \omega_I + \Aberry^{\nu, 0n}(\k) \Aberry^{\mu, n0}(\k) \omega_F}_{M^{\mu\nu}_{\text{geom}}(\k)}\right]\ND{0\k}.
\end{align}
Expanding out the Berry connection products,
\begin{align}
M^{\mu\nu}_{\text{geom}}(\k) = &-\Aberry^{\mu, 0n}(\k) \Aberry^{\nu, n0}(\k) \omega_I + \Aberry^{\nu, 0n}(\k) \Aberry^{\mu, n0}(\k) \omega_F \notag\\
=& (-i)^2 (-\braket{u_{0\k}}{\partial_{k_\mu} u_{m\k}}\braket{u_{m\k}}{\partial_{k_\nu} u_{0\k}} \omega_I + \braket{u_{0\k}}{\partial_{k_\nu} u_{m\k}}\braket{u_{m\k}}{\partial_{k_\mu} u_{0\k}} \omega_F) \\
=&-\left[ \bra{\partial_{k_\mu} u_{0\k}}\left(1-\ketbra{ u_{0\k}}{u_{0\k}}\right)\ket{\partial_{k_\nu} u_{0\k}} \omega_I 
- \bra{\partial_{k_\nu} u_{0\k}}\left(1-\ketbra{ u_{0\k}}{u_{0\k}}\right)\ket{\partial_{k_\mu} u_{0\k}} \omega_F \right]\label{eq:singlebandprojector}\\
=&- \braket{\partial_{k_\mu}u_{0\k}}{\partial_{k_\nu}u_{0\k}}\omega_I 
+ \braket{\partial_{k_\nu}u_{0\k}}{\partial_{k_\mu}u_{0\k}}\omega_F +\Aberry^\mu(\k)\Aberry^\nu(\k) (\omega_I - \omega_F),
\end{align} 
where in Eq.~(\ref{eq:singlebandprojector})  we applied the basis completeness relationship  $\sum_{m} \ketbra{u_{m\k}}{u_{m\k}} = 1 - \ketbra{u_{0\k}}{u_{0\k}} $ Decomposing the expression into symmetric and antisymmetric parts under $\mu\leftrightarrow\nu$,
\begin{align}
&-\braket{\partial_{k_\mu}u_{0\k}}{\partial_{k_\nu}u_{0\k}}\omega_I 
+ \braket{\partial_{k_\nu}u_{0\k}}{\partial_{k_\mu}u_{0\k}}\omega_F +\Aberry^\mu(\k)\Aberry^\nu(\k) (\omega_I - \omega_F)\notag \\
=&
-\frac{\omega_I + \omega_F}{2}\left[\braket{\partial_{k_\mu}u_{0\k}}{\partial_{k_\nu}u_{0\k}} - \braket{\partial_{k_\nu}u_{0\k}}{\partial_{k_\mu}u_{0\k}}\right] - \frac{\omega_I - \omega_F}{2}\left[\braket{\partial_{k_\mu}u_{0\k}}{\partial_{k_\nu}u_{0\k}} + \braket{\partial_{k_\nu}u_{0\k}}{\partial_{k_\mu}u_{0\k}}  -2 \Aberry^\mu(\k)\Aberry^\nu(\k) \right].
\end{align}
Identifying the antisymmetric part as the Berry curvature and the symmetric part as the Fubini-Study metric, we get
\begin{align}
M^{\mu\nu}_{\text{geom}}(\k) = - i \frac{\omega_I + \omega_F}{2} F^{\mu\nu}(\k) - (\omega_I - \omega_F) g^{\mu\nu}(\k).
\end{align}

\subsection{Multiband generalization via the non-Abelian QGT}

The multiband generalization follows the same algebra as above with band indices $g_1$, $g_2$ replacing $0$:

\begin{align}
    M^{\mu\nu}_{\text{geom},g_1 g_2}(\k) = -\left[\Aberry^{\mu, g_1n}(\k) \Aberry^{\nu, ng_2}(\k) \omega_I - \Aberry^{\nu, g_1n}(\k) \Aberry^{\mu, ng_2}(\k) \omega_F \right].
\end{align}
This can be simplified using the non-Abelian quantum geometric tensor, defined as $Q^{\mu\nu}_{g_1 g_2} = \braOPket{\partial_{k_\mu} u_{g_1 \k}}{(1-P)}{\partial_{k_\nu} u_{g_2 \k}}$ where $P$ is the band projector over the active manifold. By generalizing the completeness relationship in Eq.~(\ref{eq:singlebandprojector}) with the multiband version, one observes that 

\begin{align}
    \Aberry^{\mu, g_1 m}(\k) \Aberry^{\nu, mg_2}(\k) = \braOPket{\partial_{k_\mu}u_{g_1\k}}{(1-P)}{\partial_{k_\nu}u_{g_2\k}}  = Q^{\mu\nu}_{g_1 g_2}(\k).
\end{align}
It follows that 
\begin{align}
    M^{\mu\nu}_{\text{geom},g_1 g_2}(\k) = -\left[\Aberry^{\mu, g_1 n}(\k) \Aberry^{\nu, ng_2}(\k) \omega_I - \Aberry^{\nu, g_1 n}(\k) \Aberry^{\mu, ng_2}(\k) \omega_F \right]  = -Q^{\mu\nu}_{g_1 g_2}(\k)\omega_I  + \omega_F Q^{\nu\mu}_{g_1 g_2}(\k).
\end{align}
After symmetrization, one recovers Eq.~(\ref{eq:geomvertex_multiband}).
\begin{align}
    M^{\mu\nu}_{\text{geom},g_1 g_2}(\k) = - i \frac{\omega_I + \omega_F}{2} F^{\mu\nu}_{g_1 g_2}(\k) - (\omega_I - \omega_F) g^{\mu\nu}_{g_1 g_2}(\k)
\end{align}

\section{Analytical expression of effective resonant current}\label{appendix:4thorderterm}

The effective resonant scattering current in Eq. (\ref{eq:scatteringcurrento3}) originates from a double scattering process. Unlike the non-resonant term, $\hat{J}_{\text{pert}}$ admits a closed-form expression entirely within the active-band manifold, with all remote-band information encoded through the Berry connection and wavefunction overlaps. We start with the expression in the main text:
\begin{align}
	\hat{J}_{\text{pert}}(\omega_1, \omega_2)& = \sum_{m} \Proj \Big[ \Jsc{\mu}{0}{m}\frac{1}{E_0 + \omega_1 - \Ham} \hat{V}_{m0 \gets 00}~+~ \hat{V}_{00 \gets 0m} \frac{1}{E_0 + \omega_1 + \omega_2 - \Ham} \Jsc{\mu}{m}{0} \Big] \Proj.
\end{align}
In the large band gap limit $\Delta \gg \omega$, $\frac{1}{E_0 + \omega - \Ham} \sim -1/\Delta$ no longer depends on the photon frequency. Substituting the expressions of $\Jsc{\mu}{0}{m}$ and $\hat{V}_{m0\leftarrow 00}$, it becomes
\begin{equation}
\begin{split}
    \hat{J}_{\text{pert}}  = & i\sum_{m\neq 0} \sum_{\k \k'\q } \Big[\Delta_{0, m}(\k+\q) \mathcal{A}_{\k+\q}^{\mu,0m}  \frac{1}{\Delta_{0,m}(\k+\q)} U(\q) \braket{u_{m,\k+\q}}{u_{0,\k}} \braket{u_{0,\k'-\q}}{u_{0,\k'}} \\
    &+ \Delta_{m,0}(\k) \mathcal{A}_{\k}^{\mu,m0}  \frac{1}{\Delta_{0,m}(\k)} U(\q) \braket{u_{0,\k+\q}}{u_{m,\k}} \braket{u_{0,\k'-\q}}{u_{0,\k'}}\Big] \CD{0,\k+\q} \CD{0,\k'-\q} \C{0,\k'} \C{0,\k} \\
    &= i\sum_{m \neq 0} \sum_{\k \k'\q } U(\q) \Big[\mathcal{A}_{\k+\q}^{\mu,0m}  \braket{u_{m,\k+\q}}{u_{0,\k}} \braket{u_{0,\k'-\q}}{u_{0,\k'}} \\
    &-\mathcal{A}_{\k}^{\mu,m0} \braket{u_{0,\k+\q}}{u_{m,\k}} \braket{u_{0,\k'-\q}}{u_{0,\k'}}\Big] \CD{0,\k+\q} \CD{0,\k'-\q} \C{0,\k'} \C{0,\k} ,
\end{split}
\end{equation}
which depends purely on the Bloch wavefunction. One can further eliminate the coefficient via the completeness relations $\sum_{m\neq 0} \ketbra{u_{m,\k}}{u_{m,\k}} = 1 - \ketbra{u_{0,\k}}{u_{0,\k}} $  which leads to
\begin{align}
    \hat{J}_{\text{pert}}  = \sum_{\k \k'\q }U(\q)j_{\k\k'\q}^\mu\CD{0,\k+\q} \CD{0,\k'-\q} \C{0,\k'} \C{0,\k},
\end{align}
where
\begin{align}
	&j_{\k\k'\q}^\mu \\
    =& - \sum_{m \neq 0} \left( \braket{u_{0,\k+\q}}{\partial_{k_\mu} u_{m,\k+\q}} \braket{u_{m,\k+\q}}{u_{0,\k}} - \braket{u_{0,\k+\q}}{u_{m,\k}} \braket{u_{m,\k}}{\partial_{k_\mu} u_{0,\k}} \right) \braket{u_{0,\k'-\q}}{u_{0,\k'}} \notag\\
	=&- \left[ -\bra{\partial_{k_\mu} u_{0,\k+\q}} ( \sum_{m \neq 0} \ketbra{u_{m,\k+\q}}{u_{m,\k+\q}} ) \ket{u_{0,\k}} - \bra{u_{0,\k+\q}} ( \sum_{m \neq 0} \ketbra{u_{m,\k}}{u_{m, \k}}) \ket{\partial_{k_\mu} u_{0,\k}} \right] \braket{u_{0,\k'-\q}}{u_{0,\k'}} \notag\\
	=& \left[ \bra{\partial_{k_\mu} u_{0,\k+\q}} \left( 1 - \ketbra{u_{0,\k+\q}}{u_{0,\k+\q}} \right) \ket{u_{0,\k}} + \bra{u_{0,\k+\q}} \left( 1 - \ketbra{u_{0,\k}}{u_{0,\k}} \right) \ket{\partial_{k_\mu} u_{0,\k}} \right] \braket{u_{0,\k'-\q}}{u_{0,\k'}} \notag\\
	=& \left[ \braket{\partial_{k_\mu} u_{0,\k+\q}}{u_{0,\k}} + \braket{u_{0,\k+\q}}{\partial_{k_\mu} u_{0,\k}} + i ( \mathcal{A}_{\k+\q}^\mu - \mathcal{A}_{\k}^\mu ) \braket{u_{0,\k+\q}}{u_{0,\k}}  \right] \braket{u_{0,\k'-\q}}{u_{0,\k'}}.
\end{align}
Upon symmetrizing the $\k$ and $\k'$ terms, this is equivalent to the covariant derivative form used in the main text:
\begin{align}
    j_{\k\k'\q}^\mu = \frac{1}{2}(\mathcal{D}^\mu_{\k} +   \mathcal{D}^\mu_{\k'})\rho_{\k+\q, \k} \rho_{\k' -\q, \k'}.
\end{align}

\section{Analytical expressions of effective non-resonant terms}\label{appendix:nonresonantandresonanttermderivation}

In the main text, we showed that in the projected basis, the non-resonant term Raman scattering current can be decomposed into multiple non-resonant contributions. Below, we elaborate on their forms in more detail.

\subsection{Third-order term (non-resonant)}\label{appendix:3rdorderterm}

The third order term consists of scattering processes mediated with an intermediate interband coupling. Below, we represent this in the Bloch wavefunction basis and show it reduces to a Coulomb scattering process on the ground state manifold with remote-band dependent coefficients. Its explicit form is given by
\begin{align}\label{eq:appendix3rdorderterm}
\hat{\gamma}^{\mu\nu}_{\text{NR, eff}} = 
\sum_{m_1 m_2 \neq 0}  \Big[ &\Jsc{\mu}{0}{m_1} \frac{1}{E_0 + \omega_I - \Hamproj} \Jsc{\nu}{0}{m_2} \frac{1}{E_0 - \Hamproj} \hat{V}_{m_1m_2 \gets 00} \notag  \\
&+ \hat{V}_{00 \gets m_1m_2} \frac{1}{E_0 + \omega_I - \omega_F - \Hamproj} \Jsc{\mu}{m_1}{0}\!  \frac{1}{E_0 + \omega_I - \Hamproj}\!\Jsc{\nu}{m_2}{0} \notag  \\
& + \hat{\Gamma}_{0 \gets m_1}^\mu \frac{1}{E_0 + \omega_I - \Hamproj}\hat{V}_{m_10 \gets m_20} \frac{1}{E_0 + \omega_I - \Hamproj} \hat{\Gamma}_{m_2 \gets 0}^\nu  + (\mu\leftrightarrow \nu) \Big].
\end{align}
This process represents an effective non-resonant term over the ground state manifold, as the dependence on the photon frequency is suppressed at the infinite band gap limit. However, generically, this term depends explicitly on the remote band geometry and cannot be represented in terms of only Bloch wavefunctions of the active band manifold. For example, consider the first term $\Jsc{\mu}{0}{m} \frac{1}{E_0 + \omega_I - \Hamproj} \Jsc{\nu}{0}{m} \frac{1}{E_0 - \Hamproj} \hat{V}_{mm \gets 00} $. Letting $\Delta_{0, m}(\k) = E_m(\k) - E_0(\k)$ and setting $\omega \ll \Delta$, this term can be rewritten under second quantization as
\begin{align}
&\Jsc{\mu}{0}{m_1} \frac{1}{E_0 + \omega_I - \Hamproj} \Jsc{\nu}{0}{m_2} \frac{1}{E_0 - \Hamproj} \hat{V}_{m_1m_2 \gets 00} \notag \\
= &\frac{U}{L} \sum_{\p_1 \p_2 \k_1 \k_2 \k_3 \k_4} \braket{u_{0\p_1}}{\partial_{p_\mu} u_{m_1 \p_1}} \braket{u_{0\p_2}}{\partial_{p_\nu} u_{m_2 \p_2}} \braket{u_{m_1\k_1}}{ u_{0 \k_4}}\braket{u_{m_2\k_2}}{ u_{0 \k_3}}  \notag \\
&\times \frac{\Delta_{0, m_1}(\p_1)\Delta_{0, m_2}(\p_2)}{\Delta_{0, m_1}(\p_1)(\Delta_{0, m_1}(\p_1)+\Delta_{0, m_2}(\p_2))}\CD{0\p_1}\C{m_1 \p_1}\CD{0\p_2}\C{m_2 \p_2}\CD{m_1 \k_1}\CD{m_2 \k_2} \C{0 \k_3}\C{0 \k_4}  
\delta(\k_1 + \k_2 - \k_3 - \k_4),
\end{align}
which is independent of $\omega_{I}$ and $\omega_{F}$ and therefore resembles non-resonant terms. The band gap coefficient 
\begin{align}
    \frac{\Delta_{0, m_1}(\p_1)\Delta_{0, m_2}(\p_2)}{\Delta_{0, m_1}(\p_1)(\Delta_{0, m_1}(\p_1)+\Delta_{0, m_2}(\p_2))}
\end{align}
is momentum-dependent, but under a large gap limit, scales as $O(\Delta^0)$. After projecting onto the active band manifold, the sequence of fermionic operators allow a direct channel and exchange channel in terms of Wick contraction:
\begin{align}
\mathcal{P}\CD{0\p_1}\C{m_1 \p_1}\CD{0\p_2}\C{m_2 \p_2} \CD{m_1 \k_1}\CD{m_2 \k_2} \C{0 \k_3}\C{0 \k_4} \mathcal{P} = \CD{0 \k_1}\CD{0 \k_2} \C{0 \k_3}\C{0 \k_4}\left(\delta_{\p_1 \k_1} \delta_{\p_2 \k_2}- \delta_{\p_1 \k_2} \delta_{\p_2 \k_1} \delta_{m_1 m_2}\right).
\end{align}
The resulting term represents a Coulomb scattering process on the ground state manifold. However, the second term represents an exchange channel  and requires summation over all excited states $m_1 = m_2$.  The process explicitly depends on the wavefunctions of remote bands and cannot be reduced to a compact closed form in terms of the active-band wavefunctions alone. Momentum conservation imposes $\k_3 = \k_2 - \q$ and $\k_4 = \k_1 + \q$. The resulting expression has the form 

\begin{align}
    &\sum_{m_1 m_2} \Jsc{\mu}{0}{m_1} \frac{1}{E_0 + \omega_I - \Hamproj} \Jsc{\nu}{0}{m_2} \frac{1}{E_0 - \Hamproj} \hat{V}_{m_1m_2 \gets 00} = 
    \frac{U}{L} \sum_{\k_1 \k_2 \q} \CD{0 \k_1}\CD{0 \k_2} \C{0 \k_2-\q}\C{0 \k_1+\q} \notag \\
    &\Bigg( \sum_{m_1 m_2}\frac{\Delta_{0, m_1}(\k_1)\Delta_{0, m_2}(\k_2)}{\Delta_{0, m_1}(\k_1)(\Delta_{0, m_1}(\k_1)+\Delta_{0, m_2}(\k_2))}  \braket{u_{0\k_1}}{\partial_{k_\mu} u_{m_1 \k_1}} \braket{u_{0\k_2}}{\partial_{k_\nu} u_{m_2 \k_2}} \braket{u_{m_1\k_1}}{ u_{0 \k_1+\q}}\braket{u_{m_2\k_2}}{ u_{0 \k_2-\q}} \notag \\
    &-\sum_m \frac{\Delta_{0, m}(\k_1)\Delta_{0, m}(\k_2)}{\Delta_{0, m}(\k_2)(\Delta_{0, m}(\k_1)+\Delta_{0, m}(\k_2))} \braket{u_{0\k_1}}{\partial_{k_\nu} u_{m \k_1}} \braket{u_{0\k_2}}{\partial_{k_\mu} u_{m \k_2}} \braket{u_{m\k_2}}{ u_{0 \k_1+\q}}\braket{u_{m\k_1}}{ u_{0 \k_2-\q}} \Bigg)  .
\end{align}
Other terms in Eq. (\ref{eq:appendix3rdorderterm}) can be expanded in a similar manner.

\subsection{Correction of bare nonresonant term to the effective nonresonant term}\label{appendix:barenonresonantcrossterms}

Another non-resonant contribution originating from Section \ref{section:perturbationscheme} involves  cross terms between the diamagnetic vertex $\hat{\gamma}^0_{\mu\nu}$ and the interband Coulomb interaction $\hat{V}_{\text{int}}$. To distinguish the active-to-remote energy separation from the remote-band bandwidth, we write
\begin{align}
    \epsilon_m(\k)-\epsilon_g
    =
    \Delta_m+\delta\epsilon_m(\k),
\end{align}
where $\Delta_m=O(\Delta)$ and $W_r$ denotes the characteristic
scale of the dispersion of $\delta\epsilon_m(\k)$ across the Brillouin zone. Below, we assume the flat band limit $\epsilon_g = \partial_{\mu, \nu} \epsilon_g = 0$.

Expanding the bare diamagnetic vertex $\gamma^{\mu\nu}_0 $ in the band basis,
\begin{align}
    \braOPket{g}{\gamma^{\mu\nu}_0}{n} &= \braOPket{g}{\CD{g\k} u_{g o_1}(\partial_\mu \partial_\nu u^*_{o_1 m} \epsilon_m u_{o_2 m}) u_{o_2 n} \C{nk}}{n}.
\end{align}
To simplify this term, we follow the same argument as in the expansion of $\gamma^{\mu\nu}_0 $ in ground states, by noting that any term relating $\bra{g} \CD{g\k} u_{g o_1} u_{o_1 m} = \bra{g} \CD{g\k}\delta_{gm}$ must be zero in the flat band limit. The remaining non-zero term has the form
\begin{align}
    \braOPket{g}{\gamma^{\mu\nu}_0}{n} 
    &= \braOPket{g}{\CD{g\k} u_{g o_1} \partial_\mu u^*_{o_1 m}  [\partial_\nu (\epsilon_m u_{o_2 m})] u_{o_2 n} \C{nk}}{n} \\
    &= \braOPket{g}{\CD{g\k} u_{g o_1} \partial_\mu u^*_{o_1 m}  (\partial_\nu \epsilon_m u_{o_2 m} + \epsilon_m \partial_\nu u_{o_2 m}) u_{o_2 n} \C{nk}}{n} \\
    &= \braOPket{g}{\CD{g\k} (i\mathcal{A}^{\mu, g n}\partial_\nu  \epsilon_n + \sum_m i\mathcal{A}^{\mu, g m} \epsilon_m i\mathcal{A}^{\nu, mn}) \C{nk}}{n}
\end{align}
At fixed Bloch geometry and interaction, the first term is controlled by derivatives of the remote-band dispersion and scales as $O(W_r)$. The second term is weighted by active-to-intermediate-band energy differences and scales as $O(\Delta+W_r)$. Consequently, $[\gamma^{\mu\nu}_0]_{gn}=O(\Delta+W_r)$.

When this interband matrix element enters a second-order process involving one Coulomb vertex $\hat{V}_{\text{interband}}$, the resulting contribution to the non-resonant sector takes the form
\begin{align}
    \hat{\gamma}^{\mu\nu}_{\text{NR, corr}}  &= [\gamma^{\mu\nu}_0]_{gn}\frac{1}{E_0 - \Hamproj}
    \hat{V}_{n0\leftarrow 00} 
    + \hat{V}_{00\leftarrow n0}\frac{1}{E_0 + \omega_I -\omega_F - \Hamproj}[\gamma^{\mu\nu}_0]_{ng} + (\mu\leftrightarrow \nu). 
\end{align}
Holding the interband interaction strength fixed, the remote-sector resolvent scales as $O(\Delta^{-1})$, so the $\Delta-$-dependence of this correction scales as
\begin{align}
    O\left(
      \frac{\Delta+W_r}{\Delta}
    \right)
    =
    O\left[
      \left(1+\frac{W_r}{\Delta}\right)
    \right].
\end{align}
It therefore remains finite at $O(\Delta^0)$. In the model of Section~\ref{section:twoorbitalflatbandmodel}, the remote-band bandwidth itself also satisfies $W_r=O(\Delta)$.

\section{Evaluation of the resonant Raman scattering response in exact diagonalization}\label{appendix:exactdiagonalizationresponse}

Here, we describe the numerical evaluation of the resonant Raman scattering matrix in exact diagonalization. While the non-resonant contribution can be evaluated directly from the corresponding operator matrix element, the resonant scattering matrix contains the resolvent of the many-body Hamiltonian and is therefore numerically expensive. Below, we express the resonant term in terms of a Lehmann representation, which avoids the need for explicit diagonalization.

The response for a given scattering channel requires evaluating the scattering matrix according to the generalized Fermi's golden rule in Eq. (\ref{eq:ramanscatteringcrosssection}). Consider the resonant contribution to the $A_{1g}$ channel as an example, where we explictily introduce a finite broadening factor $i\eta$ for numerics:
\begin{equation}
    \begin{split}
    &\left|\braOPket{n}{\hat{M}^{xx} + \hat{M}^{yy}}{gs}\right|^2  \\
    =& \left|\braOPket{n}{J^x \left(\frac{1}{E_0 + \omega_I - H + i\eta} +\frac{1}{E_0 -\omega_F - H + i\eta}  \right)J^x + J^y \left(\frac{1}{E_0 + \omega_I - H + i\eta} +\frac{1}{E_0 -\omega_F - H + i\eta}  \right) J^y}{gs}\right|^2 .
    \end{split}
\end{equation}
To avoid explicit matrix inversion, we rewrite the sum over final states in terms of a single vector:
 \begin{equation}
     \left|\phi_{A1g}\right> =  \left[J^x \left(\frac{1}{E_0 + \omega_I - H + i\eta} +\frac{1}{E_0 -\omega_F - H + i\eta}  \right)J^x + J^y \left(\frac{1}{E_0 + \omega_I - H + i\eta} +\frac{1}{E_0 -\omega_F - H + i\eta}  \right) J^y\right] \left| gs\right>.
 \end{equation}
Similarly, the resonant contribution to the $A_{2g}$ channel, $\hat{M}^{xy}-\hat{M}^{yx}$, scatters the ground state to
 \begin{equation}
    \left|\phi_{A_{2g}}\right> = \left[J^x \left(\frac{1}{E_0 + \omega_I - H + i\eta} - \frac{1}{E_0 - \omega_F - H + i\eta}\right) J^y - J^y \left(\frac{1}{E_0 + \omega_I - H + i\eta} - \frac{1}{E_0 - \omega_F - H + i\eta}\right) J^x\right] \left| gs\right>.
\end{equation}
We can now reduce the sum over final states as
\begin{equation}
\begin{split}
    \sum_n \left|\braOPket{n}{\hat{M}^{xx} + \hat{M}^{yy}}{gs}\right|^2\delta(\Omega + E_0 - E_n) &=  \braOPket{\phi_{A_{1g}}}{\delta(\Omega + E_0 - H)}{\phi_{A_{1g}}}\\
    &=-\frac{1}{\pi} \text{Im} \left<\phi_{A_{1g}}\middle|\frac{1}{\Omega + E_0 - H + i \eta}\middle|\phi_{A_{1g}}\right>,
\end{split}
\end{equation}
which can now be evaluated as a Lehmann representation without constructing all individual eigenstates.

In the numerical evaluation, the action of each resolvent on $J^\mu \ket{gs}$ is obtained using BiCGSTAB \cite{sleijpenBiCGstabLinearEquations1993} as the solution of the corresponding linear system, e.g. $(E_g + \omega_I - H + i\eta)\ket{\alpha}= J^x\ket{gs}$, which avoids explicit inversion of the Hamiltonian.

\section{Remote-band leakage to the two-orbital flat band model}\label{appendix:projectionontogsm}

In the flat-band model calculation of Section \ref{section:twoorbitalflatbandmodel}, we performed a finite-size calculation where the discrete spectral lines are represented using Lorentzian broadening $\eta$. This broadening introduces Lorentzian tails from remote bands that extend to all frequencies and produces artificial low-energy weight from the remote-sector peaks, which would not exist in the exact spectrum. This appendix explains the origin of this artifact and the procedure used for the correction.

The full resonant Raman operator $\hat{M}_R$ acts on the ground state $\ket{gs}$ to produce a vector with components in both the active-band and remote-band sectors of the many-body Hilbert space. Schematically,
\begin{align}
    \hat{M}_R \ket{gs} = \ket{\phi} = \ket{\phi_{\text{active}}} + \ket{\phi_{\text{remote}}},
\end{align}
where $\mathcal{P}\ket{\phi_{\text{active}}} = \ket{\phi_{\text{active}}}$ and $\mathcal{P}\ket{\phi_{\text{remote}}} = 0$, with $\mathcal{P}$ the many-body projector onto states where all electrons occupy the active band.

In this model, matrix elements of the remote-sector component $\ket{\phi_{\text{remote}}}$ can scale as $O(\Delta)$. Consider the matrix element $\braOPket{n}{\hat{M}_R}{gs}$ for a final state $\ket{n}$ containing an electron in the remote band. A representative leading contribution involves an interband current vertex scattering an electron to the remote band, propagation through a virtual intermediate state, and a subsequent intraband current process within the remote band:
\begin{align}\label{eq:remoteleakage}
    \braOPket{n}{\hat{M}_R}{gs} \sim  J^{\text{remote}}_{\text{intra}} \frac{1}{\omega - H + E_0}  J_{\text{inter}} \sim O(\Delta).
\end{align}
Here, $J_{\text{inter}} \sim O(\Delta)$, the remote-sector resolvent $(\omega - H + E_0)^{-1} \sim O(\Delta^{-1})$, and $J^{\text{remote}}_{\text{intra}} = \sum_\k (\partial_{k_\mu} \epsilon_{e\k})\, \hat{c}^\dagger_{e\k}\hat{c}_{e\k} \sim O(\Delta)$, because  $\partial_{k_\mu}\epsilon_{e\k}=O(\Delta)$ as a consequence of
the $O(\Delta)$ remote-band bandwidth.

This $O(\Delta)$ matrix element produces remote-band excitation peaks of spectral weight $\sim\Delta^2$, centered at $\Omega \sim \Delta$, whose Lorentzian tails extend to the low-energy region $\Omega \ll \Delta$, suppressed by the broadening factor $\eta/\Delta^2$. At fixed $\eta$, the resulting leakage response is $\Delta$-independent at leading order:
\begin{align}
    I_{\text{leak}}(\Omega\ll\Delta) = -\frac{1}{\pi}\,\operatorname{Im}\braOPket{\phi_{\text{remote}}}{\frac{1}{\Omega + E_0 - H + i\eta}}{\phi_{\text{remote}}}  \sim O(\eta),
\end{align}
contaminating the physically relevant active-band response. To isolate the active-band response, we project out the remote-band component before computing the spectral function: $\ket{\phi} \to \mathcal{P}\ket{\phi}$. This removes the spurious leakage while preserving all contributions captured by the projected theory of Section~\ref{section:perturbationscheme}.

We emphasize that the leakage problem described above arises from the finite Lorentzian broadening applied to the finite-size exact diagonalization benchmark, vanishing in the proper limit of spectral resolution where $\eta\rightarrow 0$. Therefore, the projection $\ket{\phi} \to \mathcal{P}\ket{\phi}$ should be understood as a numerical technique to remove this contribution.

\bibliography{references}

\end{document}